\documentclass[12pt]{JHEP3}
\usepackage{epsfig}
\usepackage{amssymb}
\usepackage{bm}
\usepackage{amsmath}
\newcommand{\p}{\partial}
\def\beq{\begin{eqnarray}}
\def\eeq{\end{eqnarray}}

\newcommand{\be}{\begin{equation}}
\newcommand{\ee}{\end{equation}}
\newcommand{\cN}{{\cal N}}
\def\cN{{\cal N}}

\def\ap{{\alpha'}}

\newcommand{\Tr}{{\rm Tr}}
\newcommand{\pd}{\partial}

\makeatletter
\@addtoreset{equation}{section}
\makeatother

\newcommand{\bea}{\begin{eqnarray}}
\newcommand{\eea}{\end{eqnarray}}
\newcommand{\wt}{\widetilde}

\newcommand{\bd}{{\bf{d}}}
\newcommand{\bc}{{\bf{c}}}

\def\tr{\mathop{\rm tr}\nolimits}
\def\Tr{\mathop{\rm Tr}\nolimits}

\newcommand{\cL}{{\mathcal L}}

\newcommand{\cO}{{\mathcal O}}

\def\matt[#1,#2,#3,#4]{\left(%
\begin{array}{cc} #1 & #2 \\ #3 & #4 \end{array} \right)}

\title{Towards a Holographic Model of Color-Flavor Locking Phase}
\author{
Heng-Yu Chen$^{*,a}$, 
Koji Hashimoto$^{\dagger,b}$ and 
Shunji Matsuura$^{\S,c}$\\
${^*}$ 
{\it Department of Physics, University of Wisconsin, Madison, WI 53706,
USA}
\\
${}^\dagger$ 
{\it Theoretical Physics Laboratory, RIKEN, 
Saitama 351-0198, Japan}\\
${}^\S$
{\it Kavli Institute for Theoretical Physics, University of California,}
\\ {\it Santa Barbara CA 93106-9530, USA}\\
$^a$ E-mail: \email{hchen46@wisc.edu}\\
$^b$ E-mail: \email{koji@riken.jp}\\
$^c$ E-mail: \email{matsuura@kitp.ucsb.edu}\\
}

\abstract{We demonstrate a holographic realization of color-flavor
locking phase, using ${\cal N}=4$ $SU(N_c)$ SYM coupled to ${\cal N}=2$
$N_f$ fundamental hypermultiplets as an example. The gravity dual
consists of $N_c$ D3-branes and $N_f$ D7-branes with world volume gauge
field representing the baryon density. Treating a small number
$\tilde{N}_c \subset N_c$ of D3-branes as Yang-Mills instantons on the
D7-branes, we consider possible potential(s) on their moduli space
or equivalently the Higgs branch.  
We show that a non-trivial potential can be generated by including the
backreaction of the baryonic density on the D7-branes, 
this dynamically drives the instantons (= D3-branes) into
dissolution. We interpret this as a color-flavor locking 
since the size of the instanton is the squark vev, 
and study the symmetry breaking
patterns. Extending to finite temperature setup, we demonstrate that
color-flavor locking persists, and the thermal effect provides 
additional structures in the phase diagram. 
}

\preprint{
{\normalsize MAD-TH-09-07} \\
{\normalsize NSF-KITP-09-172} \\
{\normalsize RIKEN-TH-166}\\
{\normalsize }
}
\begin{document}
%\setcounter{page}{1}

%%%%%%%%%%%%%%%%%%%%%%%%%%%%%%%%%%%%%%%%%%%%%%%%%%%%%%%%%%%%%%
%%%%%%%%%%%%%%%%%%%%%%%%%%%%%%%%%%%%%%%%%%%%%%%%%%%%%%%%%%%%%%
%\noindent\underline{\it Introduction and summary.}
%\vspace{3mm}

\section{Introduction and Summary}
\label{sec1}
\paragraph{}
In QCD phase diagram, the color-flavor locking (CFL) phase, or more generically, the color
superconducting phase, is expected to be present in a region with
large chemical potential $\mu$ for baryon number. 
Perturbative analytic study of this phase (see Ref.~\cite{CFLReviews} for reviews) has mainly been done
for very large $\mu$ such that the QCD coupling is weak. However, the issue on
possible phase transitions from the hadronic phase at finite $\mu$ 
has not been addressed, as the system becomes strongly coupled. So far, neither direct experimental search, nor the lattice QCD simulation with ``sign problem'' have reached such region in the phase space. 

Holographic techniques from gauge/string duality \cite{Maldacena:1997re,Gubser:1998bc} may offer new insights to such issue, as they enable us to probe the strongly coupled region(s) in the phase diagram for
QCD-like theories. Although the duality strictly works for large number of colors $N_c \gg 1$,
the holographic techniques applied to QCD-like theories (so-called
``Holographic QCD'') have been rather successful in reproducing
qualitative and semi-quantitative features of low energy QCD dynamics. In this paper, among other things, we shall show that a
color-flavor locking occurs for a toy QCD-like theory at zero temperature, when the baryon chemical potential $\mu$ exceeds its critical value.\footnote{Disclaimer: Note that our theory is not QCD but rather a supersymmetric
generalization of it, and we shall only treat the squark condensation
for the CFL. For a field-theoretical treatment of the squark
condensation, see for example Ref.~\cite{Harnik:2003ke}.}
%%%%%%%%%%%%%%%%%%%%%%%%%%%%%%%%%%%%%%%%%%%%%%%%%%%%%%%%%%%%%%%%%%%%%%%%%%%%%%%

\vspace{2ex}
\noindent
\underline{Problems of CFL in Holographic QCD}
\vspace{1ex}

Before entering the details on how to realize CFL phase in our model, let us summarize here the possible difficulties in obtaining it in holographic QCD.
\begin{itemize}
\item In gauge/string duality, to treat $N_f$ flavor branes as probes, we typically need to take $N_c\gg N_f$, while the CFL refers to a locking of the $SU(3)$ flavor and the $SU(3)$ color symmetries, {\it i.e.} $N_c=N_f$.
\item In gauge/string duality, usually only gauge-invariant quantities are
considered, while in the CFL phase the order parameter is a gauge variant di-quark condensate.
\end{itemize}
%%%%%%%%%%%%%%%%%%%%%%%%%%%%%%%%%%%%%%%%%%%%%%%%%%%%%%%%%%%%%%%%%%%%%%%%%%%%%%%
The first problem is strictly technical, 
as when $N_f\sim N_c$, the backreaction of the flavor branes cannot be ignored, and renders it difficult to analyze in supergravity.\footnote{There are examples in which fully backreacted geometry is obtained (see for example Ref.~\cite{BRgeometry}), and
it would be interesting to generalize our results to those examples.}
Our approach used in this paper is to first separate some finite number
of color branes $\tilde{N}_c$({\it i.e.} 
$\tilde{N}_c \ll N_c$), and investigate the locking of 
$SU(\tilde{N}_c)$ color symmetry with the flavor symmetry. Though this procedure of
separation is artificial, our result may suggest a piece of the whole
picture. Another concern for the first problem is that in
the strict $N_c\to \infty$ limit, the theory does not reveal the CFL phase \cite{NoCFL}. We
don't consider this concern, since we will not perform a comparison with
the chiral density wave (which is supposed to be favored at the  
strict $N_c\to \infty$ limit) in our toy model, and also because 
a large but finite value of $N_c$ may give the CFL phase even for the 
real QCD \cite{MaybeCFL}.  

As for the second problem above, it is familiar to us that gauge-invariant
correlators of QCD-like theories can be computed in their gravity duals,
but in fact there are some gauge-covariant quantities which one can also
compute in the gauge/string duality.\footnote{Examples of that kind 
include computations explicitly uses string worldsheets in the dual
gravity backgrounds; gluon scattering amplitudes, drag forces,
quark-antiquark forces, heavy meson spectroscopy and Regge trajectory.}   
In this paper, we use holographic techniques for Coulomb phase of
supersymmetric Yang-Mills (SYM) 
theories \cite{Douglas:1998tk, Peet:1998wn}, where a part of the gauge
symmetry 
decouples from the rest. When the rank of this decoupled 
gauge subgroup is small, we may treat them in the same way as the probe flavor
branes, and their gauge symmetry is manifest in the dual
gravity description. We shall describe this in detail later.

\vspace{2ex}
\noindent
\underline{Supersymmetric QCD, the Holographic Dual and Phase Diagram}
\vspace{1ex}

The toy model we shall focus on is ${\cal N}=4$ SYM coupled to ${\cal N}=2$
fundamental matter hyper multiplets. The holographic dual of this theory
was proposed by Karch and Katz \cite{KarchKatz}, as a minimal deformation of the
${\cal N}=4$ SYM 
to include
quarks. The quark superfields are introduced as the lowest excitation
on a string connecting $N_c$ D3-(color-)branes and $N_f$
D7-(flavor-)branes. For $N_c\gg N_f$, D3-branes can be replaced by
$AdS_5\times S^5$ geometry, and the flavor
dynamics of the strongly coupled large $N_c$ SQCD can be analyzed by the
probe flavor D7-branes in that geometry. The quark mass $m$
quark is proportional to the distance between the
D3-branes and the D7-branes.

\begin{figure}[t]
\begin{center}
\includegraphics[width=0.4\textwidth]{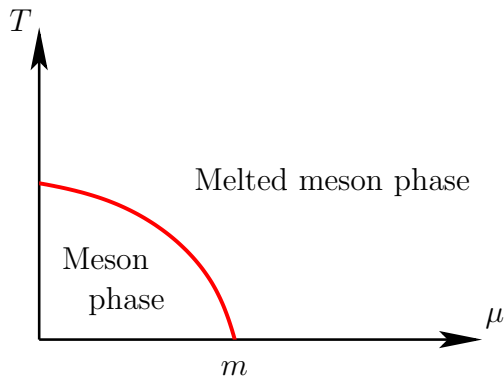}
\put(0,10){$\mu$}
\put(-180,120){$T$}
\put(-100,-10){$m$}
\put(-160,30){Meson}
\put(-150,15){phase}
\put(-110,60){Melted meson phase}
\caption{The structure of the phase diagram of the ${\cal N}=2$ SQCD.
(Scales in this figure is not
 accurate. See
 \cite{HoloBaryon,Karch:2007br,D3D7phase,Nakamura:2006xk,Erdmenger:2007ja} for details.) 
}
\label{figphase1}
\end{center}
\end{figure}

\begin{figure}[t]
\begin{center}
\includegraphics[width=0.7\textwidth]{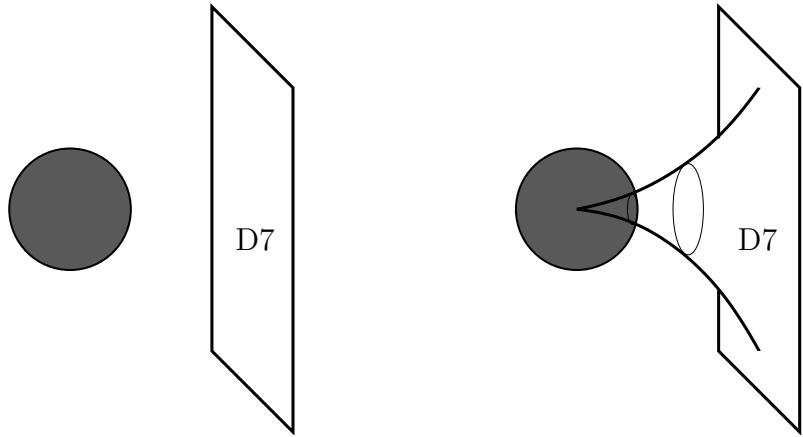}
\put(-215,70){D7}
\put(-25,70){D7}
\caption{Two embeddings of the D7-branes in the geometry. The shaded
 ball denotes a black hole with a horizon of the topology $S^5$. Left: 
Minkowski embedding (meson phase). Right: black hole embedding (melted
 meson phase). 
}
\label{figembed}
\end{center}
\end{figure}

For zero temperature $T=0$ and $\mu=0$, quarks and gluons are deconfined while 
quarks can form deeply bound mesons. 
The phase structure of this theory has been analysed by
the holographic duality \cite{HoloBaryon,Karch:2007br,D3D7phase,Nakamura:2006xk,Erdmenger:2007ja}
%(see Refs.~\cite{Kim:2006gp,} for earlier related papers), 
and at the leading large $N_c$
expansion, it is known that there are two phases in the $(\mu,T)$
diagram: the meson phase and the melted meson phase (see
Fig.~\ref{figphase1}).  
%
%\if0
%\fi
%
In both phases,
gluons are deconfined. In the meson phase, quarks are bound to form
mesons with their discrete spectrum,\footnote{The meson spectrum at zero
baryon density is studied in
Refs.~\cite{Kruczenski:2003be,Myers:2006qr}.}  
while in the melted meson phase,
the meson spectrum is continuous, and there appears nonzero 
baryon number density.
These two phases are characterized by
the shape of the probe $N_f$ D7-branes
\cite{Babington:2003vm,Kruczenski:2003uq,Kirsch:2004km,Mateos:2006nu,Albash:2006ew,Aharony:2006da}. For the finite
temperature, 
the background geometry is an AdS black hole. The meson phase
corresponds to the D7-branes away from the horizon, which is called 
{\it ``Minkowski embedding''}. On the other hand, in the melted
meson phase, the D7-branes touch the horizon (see
Fig.~\ref{figembed}), and is called {\it ``black hole embedding''}.

%
%\if0
%\fi
%

Since the local gauge symmetry $U(N_f)$ on the D7-brane is identified as
the global $U(N_f)_{\rm V}$ symmetry of the SQCD via the gauge/string
duality, the chemical potential $\mu$ is identified as the value of the temporal component of the overall $U(1)$ gauge field on the coincident
D7-branes. 
In the meson phase, this gauge field is just 
a constant $\mu$, while in the
melted meson phase, there appears electric flux on the D7-branes: 
this configuration has a lower free energy which the holographic
dual can compute, and thus favored. The phase transition is  first
order, and the critical chemical potential for $T=0$ is 
$\mu_{\rm cr}=m$. 
In the melted meson phase, 
the shape of the D7-branes is a spike whose tip is inside the
horizon. Electrically charged spikes on D-branes are 
identified as fundamental strings, % dissolved in the D-branes,
so the existence of the electric flux means that the 
quark number density is nonzero. These are briefly reviewed in
Sec.~\ref{sec2.1}. 

\vspace{2ex}
\noindent
\underline{Dynamically Driven CFL}
\vspace{1ex}

%\paragraph{}
Let us explain how the CFL phase of this theory can be realized in its gravity
dual. First of all, note that our theory is ${\cal N}=2$ SQCD which includes squarks
carrying the baryon (quark) number. So, once the chemical potential
becomes large enough, we expect squark condensation, instead of
di-quark condensation. We shall see this squark condensation in this
paper: {\it this is certainly a CFL, but also a Higgs phase}. 

As suggested before, we separate $\tilde{N}_c$ D3-branes among $N_c$ 
and treat them as probes, $\tilde{N}_c \ll N_c$. The relevant
quark/squarks are strings connecting the $\tilde{N}_c$ D3-branes
and the $N_f$ D7-branes. Condensation of strings connecting D$p$-branes
and D$(p+4)$-branes is well-known \cite{Douglas:1995bn}: 
the D$p$-branes are dissolved
into the D$(p+4)$-branes, and the D$p$-branes 
can be seen as finite size instantons on the D$(p+4)$-branes.
Therefore, the CFL Higgs phase of the SQCD is equivalent, via the
gauge/string duality, to the situation where 
the size of the instantons on the probe D7-branes is driven to become
larger. We will show in this paper that this is indeed the case, by
computing the potential of the instanton size modulus on the D7-brane
$U(N_f)$ gauge theory, in the melted meson phase.
{D3-branes are moved onto the D7-branes and dissolve on the
D7-branes dynamically.}

This Higgs phase for $T=0$ was described in Refs.~\cite{Guralnik:2004ve,
Erdmenger:2005bj, Arean:2007nh} (see also Ref.~\cite{Guralnik:2004wq}), and 
the potential for the instanton size modulus was considered in the absence of baryon density. 
At $T=0$, the resultant potential vanishes (we review it in
Sec.~\ref{sec2.2} and \ref{sec2.3}), thus there is no CFL. Our new point
is that including a backreaction from the D7-brane electric flux 
(Sec.~\ref{sec3}), this generates a nontrivial potential for the intanton size modulus (Sec.~\ref{sec4}). 
The new potential has a run-away behavior (see Fig.~\ref{figrunawayT=0}), 
causing the instantons to expand,
hence the CFL Higgs phase is prefered. 
This new potential exists only in the melted meson phase, so, for $T=0$,
above the critical baryon chemical potential, the CFL Higgs phase
appears --- this is what we show using the gauge/string duality for the
SQCD.

The way this new potential emerges is quite intriguing. This is
essentially a Chern-Simons (CS) term on the D7-branes, 
$\int {\rm tr}\; F\wedge F\wedge F \wedge C_2$. The backreaction of the
electric flux on the D7-branes generate a nonzero constant Ramond-Ramond
(RR) 3-form flux $F_3 = dC_2$ (\ref{f3back}).\footnote{The importance of
this coupling between the NSNS 2-form and the $F_3$ for the baryons
was found in
Ref.~\cite{Gross:1998gk}.} Substituting 
this to the CS 
term, we obtain $\int {\rm tr}\; A\wedge F\wedge F$, thus the electric
potential $A_t$ on the D7-brane interacts with the instanton density
${\rm tr} F\wedge F$, which gives a nontrivial potential
(\ref{potrho}).\footnote{ 
A similar CS mechanism was used for treating baryons \cite{Hata:2007mb,
Hashimoto:2009ys}
in Sakai-Sugimoto 
holographic model \cite{SaSu1}, but used in a rather different way: the CS
term was to stabilize the size of a single baryon in the model.}
This generation of $F_3$ can also be thought of being sourced by
baryon vertices, which are nothing but D5-branes wrapping $S^5$
\cite{WittenBaryon} (see also Ref.~\cite{Gross:1998gk}). 
If one smears them, they provide a constant
magnetic flux $F_3$ along $x^1$-$x^2$-$x^3$ directions
(Sec.~\ref{sec3.1}). So, our work  
is an example of backreacting baryon vertices.

\begin{figure}[t]
\begin{center}
\includegraphics[width=0.4\textwidth]{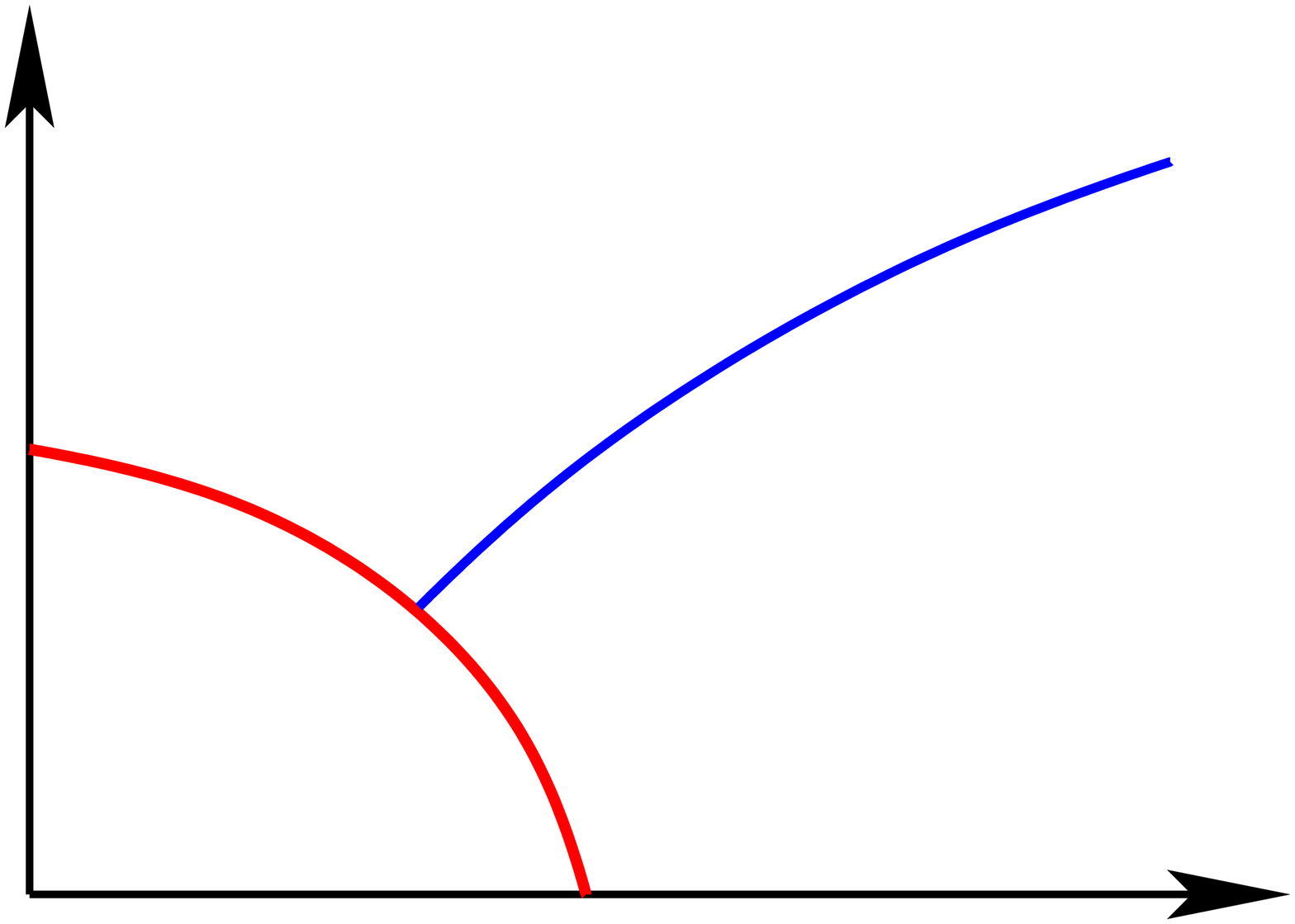}
\put(0,10){$\mu$}
\put(-180,120){$T$}
\put(-160,30){Meson}
\put(-150,15){phase}
\put(-80,40){CFL Higgs phase}
\put(-160,90){Thermal CFL}
\put(-150,75){Higgs phase}
\caption{The structure of the 
phase diagram of the ${\cal N}=2$ SQCD given by a naive
 addition of the instanton size potential coming from the back reaction
 of the geometry. The melted meson phase (corresponding to the black
 hole embedding of the D7-branes) is devided into two distinct phases.  
The upper half denoted as ``thermal CFL Higgs phase'' 
is dominated by the thermal potential of the instanton
size, in which the size 
is roughly equal to the horizon size. The lower half denoted as 
``CFL Higgs phase'' is
dominated by the one generated by the backreaction due to the baryon
density, in which the instanton size is much larger than the horizon
 size. 
}
\label{figphase2}
\end{center}
\end{figure}

We also analyze the thermalized case with $T\neq 0$
(Sec.~\ref{sec5}).  
Ref.~\cite{Apreda1} showed that, {for the finite temperature, a nontrivial potential (\ref{thermal-pot-fini-T}) 
for the size modulus of the instanton on the
D7-branes is generated, before including the baryon density. 
%our backreaction effect for the potential.
This potential is minimized at a finite value of the
instanton size. Therefore a Higgs CFL phase is prefered.  
Introducing baryonic density, we analyze the backreaction and our new CS-type potential 
(\ref{back-reac-pot-fini-T}) adds up on it. 
This addition does not change the result that the instanton size is
nonzero, so we still have the Higgs CFL phase.

If we naively adds up the two potentials (the thermal potential
(\ref{thermal-pot-fini-T}) given in Ref.~\cite{Apreda1} and 
our CS-type potential (\ref{back-reac-pot-fini-T})), 
we find that there are two
CFL phases: for small values of the baryon density, we have the
minimum at a finite value of the instanton size, while for large values
of the density the minimum sits at the infinite size modulus.
An expected phase diagram is given in Fig.~\ref{figphase2}. 
However, since the potential computed in this paper is valid only
around small size of the instanton, this conclusion is a qualitative one and 
deserves further study for full backreaction of the geometry.
}

We conclude by discussing some interesting future directions in Sec.~\ref{sec6}.
%
%\if0
%\fi
%

\section{Instanton on the Flavor Branes}
\label{sec2}
\paragraph{}
We start with constructing a solution for the equations of motion on the
$N_f$ flavor D7-branes, which has nonzero instanton number. This
solution corresponds to the dissolved D3-branes in the flavor
D7-branes. In this section, the probe approximation for the D7-branes is
adopted, while the important backreaction will be treated in
Sec.~\ref{sec3}, and its effect on the solution which we will find in
this section will be studied in Sec.~\ref{sec4} where 
the dynamical dissolution of the instantons (D3-branes) due to the
finite baryon density is shown.

\subsection{Review of the D3D7 System at Finite Baryon Density}
\label{sec2.1}
\paragraph{}
Let us begin for simplicity, by considering the case with 
zero temperature, which corresponds to $AdS_5\times S^5$ background in
type IIB Supergravity.
We shall embed in it a stack of $N_F$ space-time filling D7-branes, with a non-trivial world volume baryonic $U_b(1)$ gauge field turned on.
As it turns out, the exact shape of D7s and profile of the gauge field can be analytically solved in such regime \cite{Karch:2007br}, which we shall review in some detail next.

The $AdS_5\times S^5$ metric, as generated by the backreaction of $N_c$ D3-branes, is given in Poincare coordinates:
\begin{eqnarray}
ds^2 &=& \frac{r_6^2}{R^2}\eta_{\mu\nu}dx^\mu dx^\nu 
+\frac{R^2}{r_6^2}\left(dr_6^2 + r_6^2 ds_5^2\right)\,,~~~\frac{R^4}{\alpha'^2}=4\pi g_s N_c=\lambda \label{AdS5S5metric}\,,\\
g_s C_4&=&\frac{r_6^4}{R^4}dx^0\wedge dx^1\wedge dx^2\wedge dx^3\,, \label{C4}\\
g_s F_5&=&(1+*_{10})d(g_s C_4)=4R^4(d\Omega_5+*_{10} d\Omega_5)\,.\label{F5}
\end{eqnarray}
Here we have listed out the RR 4-form field $C_4$ and the self-dual
5-form field strength $F_5$, whereas the string coupling
$g_s=e^{\Phi_0}$ remains constant. The indices $\mu\,\nu$ runs over
$0,1,2,3$, $\eta_{\mu\nu}$ denotes four dimensional Minkowski metric,
and $ds_5^2$ is the metric for a unit five-sphere. For our later
purpose, let us also reparametrize the flat six internal dimensional
metric as: 
\begin{equation}
dr_6^2+r_6^2 ds_5^2=dr^2+r^2 ds_3^2+dy^2+dz^2\,,\label{reparaR6}
\end{equation}
with $r_6^2=r^2+y^2+z^2$. Here 
$ds_5^2$ ($ds_3^2$) is the metric on the unit
$S^5$ ($S^3$).
In such coordinates, there exists $U(1) \subset SO(6)$ isometry group which rotates $(y,z)$.    

Introducing $N_f$ $(\ll N_c)$ probe D7-branes into
(\ref{AdS5S5metric}), their Dirac-Born-Infeld (DBI) action is given by  
\cite{Myers:1999ps}
\begin{eqnarray}
S_{\rm DBI}^{\rm D7} = -{\cal T}_{\rm D7}\int d^8\xi \;
e^{-\Phi} \; {\rm tr} 
\sqrt{-\det(G_{ab} 
+ 2\pi\alpha'F_{ab})}\,.
\label{d7}
\end{eqnarray}
Here $\xi^a, a=0,\dots 7$ are the eight dimensional D7-brane world
volume coordinates, $G_{ab}$ is the pullback metric and $F_{ab}$ is the
worldvolume gauge field, which for now, we shall only turn on the
diagonal baryonic $U_b(1)$ component. ${\cal T}_{\rm D7}e^{-\Phi_0}
= 1/((2\pi)^7 \alpha'^4 g_s)$ is the tension of the D7-brane.
{The trace is taken over the symmetrized gauge indices.
Note that the symmetrized trace is valid only up to the fourth order in $\ap$
 \cite{Tseytlin:1997csa, Hashimoto:1997gm, Tseytlin:1999dj}.
However, 
it is known that, at this order, the non-abelian DBI equations are solved 
at least for the instanton configurations.}

We choose the gauge for the D7-brane worldvolume coordinates as
\begin{equation}
(\xi^0,\dots\xi^3)\equiv (t,\dots,x^3)\,,~~~(\xi^4,\dots,\xi^7)\equiv
 (r,S^3)\, , 
\end{equation}
so that the D7-branes are spacetime filling and spanning in the four flat internal directions given by $r$ and $S^3$ in (\ref{reparaR6}). The D7-branes therefore have asymptotic worldvolume geometry of $AdS_5\times S^3$. 
The precise D7-brane embedding are specified by the transverse coordinates $(y,z)$, which become D7-brane scalar fields. To preserve the isometry of $S^3$, we have $(y(r),z(r))$; the $U(1)$ isometry further sets $z(r)=0$. The induced D7-brane world volume metric is therefore:
\begin{equation}
G_{ab}d\xi^a d\xi^b=\frac{(r^2+y(r)^2)}{R^2}\left(\eta_{\mu\nu}dx^\mu dx^\nu\right)+\frac{R^2}{(r^2+y(r)^2)}\left((1+(y'(r))^2)dr^2+r^2 ds_3^2\right)\,,\label{Gab}
\end{equation} 
where $'$ denotes $\frac{d}{dr}$. Turning on only the temporal component
of the $U_b(1)$ gauge field $A_t(r)$, which we again take to be
dependent purely on $r$, the resultant D7-brane DBI action
(density)\footnote{We divide out the infinity volume of four Minkowski
spacetime $V_4$.} is: 
\begin{equation}
S_{\rm DBI}^{\rm D7}/V_4=\int dr L=-{\cN}\int dr \;
r^3\sqrt{(1+(y'(r)^2)-(2\pi\alpha' A_t'(r))^2}\label{d7action}\,,
\end{equation}   
where $\cN= N_f {\cal T}_{\rm D7} {\rm
Vol}(S^3)g_s^{-1}=N_f{\mathcal{T}}_{\rm D7}(2\pi^2)g_s^{-1}$, and factor $N_f$ arises from the trace.

As noted in Ref.~\cite{Karch:2007br}, the action (\ref{d7action}) does not contain explicit dependences on $y(r)$ and $A_t(r)$, their equations of motion yield following constants of motion:
\begin{eqnarray}
\frac{\delta L}{\delta y'}
&=&-{\mathcal{N}} r^3 \frac{y'}{\sqrt{1+(y')^2-(2\pi\alpha' A_t')^2}}
=-\bc\,,\label{constc}\\
\frac{\delta L}{\delta(2\pi\alpha' A_t')}
&=&{\mathcal{N}} r^3 \frac{2\pi\alpha'A_t'}
{\sqrt{1+(y')^2-(2\pi\alpha' A_t')^2}}=\bd\,.
\label{constd}
\end{eqnarray}
A useful relation can also be readily deduced 
\begin{equation}
2\pi\alpha' 
A_t'(r)=\frac{\bd}{\bc}y'(r)\, .\label{Usefulrel}
\end{equation}
Using this and rearranging (\ref{constc}) and (\ref{constd}), we obtain   
\begin{eqnarray}
 2\pi\alpha' A_t'(r) = \frac{\bd}{\cN\sqrt{r^6 + r_0^6}}, \quad
 y'(r) = \frac{\bc}{\cN\sqrt{r^6 +r_0^6}}\,, \label{KOsol}
\end{eqnarray}
where we have defined:
\begin{equation}
r_0^6=\frac{\bd^2-\bc^2}{\cN^2}\,.\label{Defr0}
\end{equation}
We can readily integrate (\ref{KOsol}) to obtain the profiles for $y(r)$
and $2\pi\alpha' A_t(r)$: 
\begin{eqnarray}
&&y(r)=\frac{\bc}{2~3^{1/4}\cN r_0^2}{\mathbb{F}}
\left(\!\varphi(r),\frac{2+\sqrt{3}}{4}\right)
\,,~2\pi\alpha' A_t(r) =\frac{\bd}{2~3^{1/4}\cN r_0^2}{\mathbb{F}}
\left(\!\varphi(r),\frac{2+\sqrt{3}}{4}\right)\,,\nonumber\\
\label{yrAr}\\
&&\varphi(r)=\arccos\left(\frac{1-(\sqrt{3}-1)(r/r_0)^2}{1+(\sqrt{3}+1)(r/r_0^2)^2}\right)\,,\label{varphir}
\end{eqnarray}
where ${\mathbb{F}}(\varphi,k)$ is the incomplete elliptic integral of the first kind. In the above computations, we have taken $\bd>\bc$, the resultant solutions (\ref{yrAr}) should be regarded as the zero temperature analog of the aforementioned black hole embedding \cite{HoloBaryon,Karch:2007br}. In such case, the D7-branes extend all the way to the ``horizon'' located at $\sqrt{r^2+y(r)^2}=r_6=0$ ($z(r)$ has been set to zero) and we have used this fact to fix the integration constant. 
The profile of $y(r)$ in (\ref{yrAr}) in $(r,y(r))$ plane displays a sharp peak towards $y(0)=0$ around $r=0$ (or four dimensional cone when sweeping out the $S^3$), and flattens out to approach $2\pi\alpha' m$ as $r\to \infty$, where $m$ is the bare quark mass.  
This is in contrast with the Minkowski embedding where D7-branes lie at finite distance from the horizon, or $\sqrt{r^2+y(r)^2}>r_H$.  
In the presence of finite baryon density, it was shown in Ref.~\cite{HoloBaryon} that only black hole embedding is stable and physical, we shall discuss them in more details in section \ref{sec5}.  

Finally one can relate the asymptotic values of $y(r)$ and $2\pi \alpha' A_t(r)$ with the quark mass $m$ and the chemical potential $\mu$ as $y(\infty)\to 2\pi\alpha' m$ and $2\pi \alpha' A_t(\infty)\to 2\pi\alpha' \mu$ \cite{HoloBaryon}, and obtain the following relations \cite{Karch:2007br}:    
\begin{eqnarray}
&& \bc = \gamma {\cal N} (2\pi\alpha')^3 (\mu^2-m^2)m \, ,  
\label{relm}\\
&& \bd = \gamma {\cal N} (2\pi\alpha')^3
 (\mu^2-m^2)\mu \, .
\label{relmu}
\end{eqnarray}
Here the constant $\gamma=\left(\frac{\sqrt{\pi}}{\Gamma(1/3)\Gamma(7/6)}\right)^{-3}\sim 0.363$.
This completes our review on the zero temperature D7-brane embedding in the presence of baryonic $U_b(1)$ gauge field. To realize the color-flavor locking phase, we shall next consider turning on a $SU(N_f)$ instanton configuration within the internal four cycle as a perturbation. 
%%%%%%%%%%%%%%%%%%%%%%%%%%%%%%%%%%%%%%%%%%%%%%%%%%%%%%%%%%%%%%%%%%%%%%%%%%%%%%%

\subsection{Instanton Solution}
\label{sec2.2}
\paragraph{}
We are ready to consider the non-Abelian
part of the $U(N_f)$ gauge group on the flavor D7-branes, including the
instantons. 
In the equations of motion, the overall $U_b(1)$ discussed earlier is
coupled to the $SU(N_f)$ subsector where we like to put the instantons
representing the D3-branes. 

As we shall see later, for large 'tHooft coupling $\lambda$, the
non-Abelian part can be regarded as a fluctuation around the fixed
$U_b(1)$ background (\ref{yrAr}).
We substitute the
$U_b(1)$ solution (\ref{KOsol}) of Ref.~\cite{Karch:2007br} into the action and
consider only the $SU(N_f)$ non-Abelian part of the action (\ref{d7}). 
We are interested in solutions having the instanton charges in the subspace
$(\xi^4, \cdots, \xi^7)$, so we just turn on $SU(N_f)$
$A_i(\xi)$ $(i=4, \cdots, 7)$ among the gauge fields, and let them be
dependent on only the coordinates $\xi^i$. 

In the action, the effective four cycle metric felt by these non-Abelian components is computed as follows. We note that the background $A_t(r)$ 
can be regarded as an additional transverse scalar field in the D7-brane DBI
action, as we are interested in only the space spanned by 
$(\xi^4, \cdots, \xi^7)$.
Indeed, the effective metric for the directions 
$(\xi^4, \cdots, \xi^7)$ can be written formally as
\begin{eqnarray}
 G_{ij}^{(4)} = g_{ij}
+ g_{yy} \p_i y \p_j y  + g^{tt} \p_i A_t \p_j A_t  (2\pi\alpha')^2\,,~~~i,j=4,5,6,7.\label{Gij1}
\end{eqnarray}
Since $y(r)$ and $2\pi\alpha' A_t(r)$ are functions of
$r=\sqrt{\sum_{i=4}^7 (\xi^i)^2}$, we can rewrite above as 
\begin{eqnarray}
{G}_{ij}^{(4)} = \frac{R^2}{r_6^2}
\left(\delta_{ij} + \frac{\xi^i\xi^j}{r^2}
\left(y'^2 - (2\pi\alpha' A_t')^2\right)\right).\label{Gij2}
\end{eqnarray}
So the determinant in 
the DBI action, including the non-Abelian field strength $F_{ij}$ in the
$SU(N_f)$, 
is written as
\begin{eqnarray}
-\det (G_{ab}+2\pi\alpha' F_{ab})
& = &
\det \left(
\widetilde{G}_{ij}^{(4)}
+ 2\pi\alpha' F_{ij} \frac{r^2+y^2(r)}{R^2}
\right)
\end{eqnarray}
where the unwarped effective four cycle metric $\widetilde{G}_{ij}^{(4)}$ is given by
\begin{eqnarray}
{\widetilde{G}}_{ij}^{(4)} \equiv
\delta_{ij} 
+ \frac{\xi_i\xi_j}{r^2}(y'^2 - (2\pi\alpha' A_t')^2) \, .\label{tGij}
\end{eqnarray}
Thus the total DBI action including the non-Abelian field strengths
$F_{ij}$ is 
\begin{eqnarray}
 S_{\rm DBI}^{\rm D7} = -{\cal T}_{\rm D7} 
\int\! d^4x \int\! d^4\xi
\; e^{-\Phi} \; {\rm tr} 
\sqrt{
\det \left(
\widetilde{G}_{ij}^{(4)} + 2\pi\alpha' F_{ij}\frac{r^2+y^2(r)}{R^2}
\right)} \,.
\label{dbiin}
\end{eqnarray}
In this expression, note that the prefactor of $F_{ij}$ is
suppressed by $\lambda^{-1/2}$.  In fact, 
$2\pi \alpha'/R^2 = 2\pi/\sqrt{\lambda}$, 
with the relation $R^4 = 4\pi g_s N_c \alpha'^2$. 
We can therefore regard
the instanton as a fluctuation around the fixed $U_b(1)$ background sourced by $A_t$, for a large $\lambda$.

In addition to this DBI action, now we also have a Chern-Simons (CS) term by coupling with background RR 4-form $C_4$ given in (\ref{C4}) with  
the non-Abelian field strength $F_{ij}$:
\begin{eqnarray}
S_{\rm CS}^{\rm D7} = {\cal \mu}_{\rm D7} \int \! d^4x
\int \! d^4\xi \; \frac{1}{g_s} 
\left(\frac{r^2 + y(r)^2}{R^2}\right)^2
\frac{(2\pi\alpha')^2}{8} \; {\rm tr}
\left[\epsilon^{ijkl} F_{ij}F_{kl}\right]\,,
\label{csd7}
\end{eqnarray}
with $\mu_{\rm D7}= {\cal T}_{\rm D7}$
We will show that self-dual configurations of the 
non-Abelian gauge fields with respect to the metric 
${\widetilde{G}}^{(4)}_{ij}$ satisfies a particular property:
the $F_{ij}$-dependent part of the DBI action (\ref{dbiin}) is completely
canceled by the Chern-Simons term (\ref{csd7}). This interesting
property of the instantons on the D7-branes was explicitly shown for a
special case in
Ref.~\cite{Arean:2007nh} 
which treated the case of the flat D7-branes ($\bc=\bd=0$).
We use the following formula in generic curved space
\cite{Gibbons:2000mx}
\begin{eqnarray}
 \sqrt{{\rm det} (g+F)} = \sqrt{\det g} + \frac14 \sqrt{{\rm det} g}
\left|
F_{ij} *_4 F^{ij}
\right| 
\end{eqnarray}
for the self-dual configuration 
\begin{eqnarray}
F_{ij}= *_4 F_{ij}\,. \label{selfdual}
\end{eqnarray}
Here the Hodge dual operation $*_4$ is with respect to the effective four cycle metric and defined by a covariant
totally antisymmetric tensor $\eta^{ijkl}$, 
\begin{eqnarray}
 *F^{ij}\equiv \frac12 \eta^{ijkl}F_{kl}\, , \quad \eta^{ijkl} =
  \frac{1}{\sqrt{\det g}} \epsilon^{ijkl}\, , \quad \epsilon^{4567}=1\, .
\end{eqnarray}
This formula was shown in 
Ref.~\cite{Gibbons:2000mx} for Abelian field strength, and now if
we assume that the non-Abelian DBI action is written with the symmetric
trace prescription, this equality 
also holds for the present non-Abelian case.
Once we apply this formula to our DBI action (\ref{dbiin}), 
for the self-dual instanton configuration with respect to the metric 
${\widetilde{G}}^{(4)}_{ij}$, we obtain
\begin{eqnarray}
S_{\rm DBI}^{\rm D7} = -\frac{{\cal T}_{\rm D7}}{g_s}
\int \! \!d^4x \!\!\int \!\! d^4\xi \;{\rm tr}
\left[
\sqrt{\det {\widetilde{G}}^{(4)}_{ij}}
+ \frac{(2\pi\alpha')^2}{8}
\left(\frac{r^2\! +\! y(r)^2}{R^2}\!\right)^2
\!\epsilon^{ijkl} F_{ij} F_{kl}
\right]\,.
\end{eqnarray}
Note that we rewrite the Hodge dual by the constant tensor
$\epsilon^{ijkl}$. Using the relation ${\cal T}_{\rm D7}=\mu_7$ and $g_s=e^{\Phi}$ is fixed, 
it is obvious that the $FF$ dependent terms in the DBI is canceled by the CS actions (\ref{csd7}). 

It is interesting that this cancellation occurs not only for the flat
D7-branes with no electric flux on it but also our present case, albeit
our D7-brane configuration breaks the supersymmetries completely. 
The state with the D3 branes and the D7 branes connected by the fundamental strings in flat space is supersymmetric. However, in our case, the spike does not extend to infinity, supersymmetry is thus broken.
In Ref.~\cite{Arean:2007nh}, it was argued that this cancellation is due to
the BPS property of the D3D7 system. Here we could show the same
cancellation even with the non-supersymmetric electric flux, and  
there is therefore no potential on the instanton moduli space. 

However, in the remaining part of this paper, we will see that in fact a
backreaction of this electric flux on the D7-branes will lift the
cancellation slightly, and induces a potential term for the instanton
moduli space. It is an essential point which we like to focus on in
this paper.

\subsection{Conformal Metric and Explicit Instanton Configuration}
\paragraph{}
\label{sec2.3}
Our self-dual configuration of the non-Abelian field strength is with
respect to the curved ``effective'' metric
${\widetilde{G}}^{(4)}_{ij}$. On the other hand, the simpler case of 
Ref.~\cite{Arean:2007nh} has a flat metric $\delta_{ij}$ instead. 
In the following, we show that 
a coordinate transformation can turn the effective unwarped four cycle metric 
${\widetilde{G}}^{(4)}_{ij}$ (\ref{tGij}) into a conformally flat metric, so that in the
new coordinate the standard  BPST instanton configuration suffices.
In any conformally flat space, the self-dual equation on it is 
simply the same as the self-dual equation on the flat space.

It is easy to see that the metric ${\widetilde{G}}_{ij}^{(4)}$ (\ref{tGij}) can be written in the polar coordinate as
\begin{eqnarray}
ds^2 = \left(1 + (y'^2-(2\pi\alpha' A_t')^2 )\right)dr^2 + r^2
 ds_3^2 \, .\label{z0metric} 
\end{eqnarray}
Substituting the explicit expressions for $y'(r)$ and $2\pi\alpha' A'_t(r)$ (\ref{KOsol}), we can deduce that 
\begin{equation}
1+y'^2(r)-(2\pi\alpha' A_t'(r))^2 =\frac{r^6}{r^6+r^6_0}\,.
\end{equation}
To show (\ref{z0metric}) is conformally flat, let us consider
\begin{eqnarray}
ds^2 = \left(\frac{r^6}{r^6+r_0^6}\right)dr^2 + r^2
 ds_3^2=S(\tilde{r})^2 
\left(d \tilde{r}^2 + \tilde{r}^2 ds_3^2\right)
\end{eqnarray}
and solve for $\tilde{r}$ and $S(\tilde{r})$.
First the consistency in $S^3$ directions demands that
$r=S(\tilde{r})\tilde{r}$, the relevant differential equation in the $r$ and
$\tilde{r}$ directions then gives: 
\begin{equation}
\frac{d\tilde{r}}{\tilde{r}}=\frac{r^2}{\sqrt{r^6+r_0^6}}dr\,.
\label{conftrans}  
\end{equation}
Integrating both sides, we can obtain the desired change of variable, 
\begin{equation}
\tilde{r}=r\left[\frac{1+\sqrt{1+r_0^6/r^6}}{2}\right]^{1/3}.
\label{rtrrelation1}
\end{equation}
The integration constant is fixed so that $r\sim \tilde{r}$ for large
$r$. 
We can also invert the relation (\ref{rtrrelation1}) to obtain
\begin{equation}
\frac{r}{\tilde{r}}=S(\tilde{r})=\left[1-\frac{r_0^6}{4\tilde{r}^6}\right]^{1/3}\,.
\label{rtrrelation2}
\end{equation}
In this new coordinate $\tilde{r}$, the self-dual configuration is just
the familiar BPST instanton. When bringing that to the original
coordinate $r$, we obtain a solution to the self-dual equation in the
space with the metric $\widetilde{G}_{ij}^{(4)}$.
In Sec.~\ref{sec4}, we shall use this explicit coordinate transformation to
evaluate the potential for the instanton size moduli.

\section{Linearized Supergravity Backreaction}
\label{sec3}
\paragraph{}
In this section, we shall compute a linearized perturbation to the supergravity
background (\ref{AdS5S5metric}), (\ref{C4}), (\ref{F5}), due to the
electric field $A_t$ on the D7-branes. 
Let us first recall that the electric flux, which is responsible for the
$U_b(1)$ baryon charge, can be regarded as fundamental strings dissolved
in the D7-branes. 
This is because in the DBI action the electric field is combined
with the (pull-back of) NSNS 2-form field $\hat{B}_2$ in a gauge-invariant fashon,
$2\pi\alpha' F_{ab} + \hat{B}_{ab}$.  
Such electrified D7-branes can be regarded as a source to the bulk 3-form
flux $H_3\equiv dB_2$, 
acting as small perturbation to the background SUGRA
solution. 
Moreover from the consistent equations of motion of the SUGRA, 
this also induces RR 3-form flux $F_3$, which we will proceed to
extract in two different ways. 
The induced $F_3$ is important for the dynamics of the
instantons on the D7-branes as we will see in the next section. So, in
this section, we derive the exact amount of this $F_3$ as a backreaction of the electrified D7-brane configuration, which is
\begin{eqnarray}
F_{123}^{(3)} = \frac{8\pi^3\alpha'^2 \bd}{N_c} \, .
\label{f3back}
\end{eqnarray}

First in Sec.~\ref{sec3.1}, we present an intuitive derivation of the
$F_3$ by using smeared baryon vertices. In 
Sec.~\ref{sec3.2}, 
we compute the backreaction to the geometry due to
the electrified D7-branes. The result of 
Sec.~\ref{sec3.2} coincides with that of
Sec.~\ref{sec3.1}.

\subsection{Smeared Baryon Vertices}
\label{sec3.1}

The electric fields on the D7-branes are interpreted as fundamental
strings connecting the D7-branes and the D3-branes, therefore they are
quarks. The number density of them is given by (\ref{constd}), quark
density $= 2\pi\alpha' \bd$. This means that the baryon number density
is $2\pi\alpha' \bd/N_c$.

The D7-brane spike terminates at the origin $r_6=0$. If we take the
flux conservation at the tip of the spike seriously, we need to assume
the presence of the baryon vertices surrounding the origin. As is well
known, D5-branes wrapping the $S^5$, which are called baryon vertices, 
can give a charge at which the fundamental strings can end
\cite{WittenBaryon}. 
In this subsection, we compute 
a back reaction of these baryon vertices smeared
on the plane $x^1$-$x^2$-$x^3$ at $r_6=0$. Our result is
(\ref{f3back}).\footnote{A related issue on backreaction of baryon
vertices was discussed in Ref.~\cite{Hashimoto:2008jq}.}

The relevant terms from the type IIB supergravity and D5-brane DBI
actions are 
\begin{eqnarray}
-\frac{1}{4 \kappa_{10}^2} \int \! d^{10}x \;
\sqrt{-g_{10}} |F_7|^2 + \mu_5 \int C_6 \, ,\label{D5actions}
\end{eqnarray}
with $4 \kappa_{10}^2 = 2 (2\pi)^7\alpha'^4$ and 
$\mu_5 = (2\pi)^{-5} \alpha'^{-3}$. 
We have also used $F_7=dC_6=*_{10} F_3$.
It is enough to consider the explicit component 
$C_6 = C^{(6)}_{0\theta_1\theta_2\theta_3\theta_4\theta_5}
dx^0\wedge d\theta_1\wedge d\theta_2 \wedge d\theta_3 \wedge
d\theta_4 \wedge d\theta_5$, then (\ref{D5actions}) becomes
\begin{eqnarray}
\int d^4x dr_6 d\theta_1 d\theta_2 d\theta_3 d\theta_4 d\theta_5
\left[
\frac{-1}{4\kappa_{10}^2} \frac{r_6^3}{R^8}\frac{1}
{\sin^4\theta_1\sin^3\theta_2\sin^2\theta_3\sin\theta_4}
(\partial_{r_6}C^{(6)}_{0\theta_1\theta_2\theta_3\theta_4\theta_5})^2
\right.
\nonumber 
\\
\left.
+ \mu_5 \frac{2\pi\alpha' \bd}{N_c} \delta(r_6-\epsilon) 
C^{(6)}_{0\theta_1\theta_2\theta_3\theta_4\theta_5}
\right],\qquad
\end{eqnarray}
where the position of the baryon vertices is specified as $r_6=\epsilon$
with $\epsilon \to 0$. This can be solved as
\begin{eqnarray}
\partial_{r_6}C^{(6)}_{0\theta_1\theta_2\theta_3\theta_4\theta_5}
= \frac{8\pi^3\alpha'^2 \bd}{N_c} \frac{R^8}{r_6^3}
\sin^4\theta_1\sin^3\theta_2\sin^2\theta_3\sin\theta_4 \, ,
\end{eqnarray}
where we have also used the explicit expressions for $\kappa_{10}^2$ and $\mu_5$. Taking a Hodge dual in the background $AdS_5\times S^5$, we immediately obtain (\ref{f3back}).

%%%%%
The above analysis leads to an important consequence 
which resolves a problem in introducing baryons in D$p$/D$q$ systems.
The phase structure of fundamental matter at finite baryon density has
been studied by introducing electric flux on probe D$q$-branes in
D$p$-brane background \cite{HoloBaryon,D3D7phase,Matsuura:2007zx}.
The baryon number there was considered to be carried by free quarks in
the sense that the quark density, or the electric flux, can take any
value as long as the total number of strings takes an integer. 
In other words ${\bf d}$ is quantized in units of 1.\footnote{In this
paper this ${\bf d}$ is the density, but one can imagine localized
quarks/baryons instead, for the discussion here.}
It is natural to ask what if 
one considers baryons instead of the quarks in the system. 
As was pointed out in 
Refs.~\cite{HoloBaryon,Matsuura:2007zx} and studied in detail in
Ref.~\cite{Seo:2008qc}, it turned out that 
there is no stable baryon vertex solution 
outside the horizon, in the deconfinment phases.

A resolution of this problem of the missing baryon vertex 
is that the baryon vertices undergo a brane/flux transition 
and leaves only RR flux outside the horizon. The DBI part of the D5-brane 
baryon vertex disappears since its volume element vanishes (the time
direction of the geometry shrinks), while the CS term of the D5-brane
action remains to source the bulk RR 3-form flux $F_3$. We can see this
``remnant'' of the baryon vertices anyway, by solving
consistently the SUGRA equations of motion for the NSNS $B$-field,
as in the following Sec.~\ref{sec3.2}. In this section we work with
$T=0$, but the role of the horizon is played by the origin $r_6=0$.
Similar transition can be found in a simpler example.
Consider
$AdS_5 \times S^5$ with $N$ units of the RR flux and put an additional probe D3 brane parallel to the boundary in this spacetime
at certain $r_6$.
This is a supersymmetric configuration (the Coulomb phase) and $r_6$ is a modulus.
When the brane goes to $r_6=0$, the DBI part of the D3 brane becomes zero.
The correct picture for this case is given by $AdS_5 \times S^5$ with $N+1$ units of the flux.
Therefore, the probe D3 brane is replaced by a unit of flux.
This argument  can be applied to the finite temperature case.

It is interesting that 
the quantization condition of this $F_3$ in (\ref{f3back}) shows that 
the quark number density {\bf d} 
is quantized not in units of 1 but in units of $N_c$
\footnote{The flux is smeared along the space directions parallel to the boundary: $x_1,x_2,x_3$.
With these directions being non-compact, we do not need to quantize the flux from computational point of view.
However, our motivation of this quantization comes from the fact that the flux is sourced by the D5 branes.
}. 
This would suggest that 
the quarks in the D$p$/D$q$ system are always thought to be 
components of baryons. 
We will see in Sec.~\ref{sec5} that 
the analysis of Sec.~\ref{sec3.2} 
still applies to a finite temperature system
despite the fact that  the end points of the strings are hidden inside
the black hole horizon.

%%%%%%%%

\subsection{Backreaction from the D7-brane Electric Flux}
\label{sec3.2}
\paragraph{}
Instead of assuming the presence of the D5-brane baryon vertices,
here we provide an alterative derivation for (\ref{f3back}) by solving
the backreaction due to the electric flux on the D7-branes. 
In this subsection,
we demonstrate  
this by looking at the equation of motion for the NSNS 2-form field
$B_2$. 
For the validity of our approximation adopted in this section, see
Sec.~\ref{sec323}. 

%%%%%%%%%%%%%%%%%%%%%%%%%%%%%%%%%%%%%%%%%%%%%%%%%%%%%%%%%%%%%%%%%%%%
\subsubsection{Sourcing the Bulk NS-NS B-Field}
%\paragraph{}

First, let us examine how the electric flux on the D7-branes can act as a source for the bulk NSNS 2-form field $B_{2}$. The DBI action includes 
the NSNS B-field as
\begin{eqnarray}
 S^{\rm D7}_{\rm DBI} = - {\cal T}_{\rm D7}
\int \!d^8\xi \; 
e^{-\Phi}\; {\rm tr} \sqrt{-\det (g_{ab} + 2\pi\alpha' F_{ab}
+ \hat{B}_{ab})}\,,\label{DBIBaction}
\end{eqnarray}
where $\hat{B}_{ab}$ is the induced NSNS $B$-field carried by the fundamental strings in the D7-branes.
We are treating here only the overall $U_b(1)\subset U(N_f)$ sector (\ref{KOsol})
only, so tracing over already gave the factor $N_f$. Since 
we are interested in a linear perturbation by this source, 
we expand this action around $\hat{B}=0$ to the linear order in the $B$-field:
\begin{eqnarray}
S^{\rm D7}_{\rm DBI}\biggm|_{{\cal O}(\hat{B})}
= -\int d^4x dr
\; \hat{B}_{0r}
\left[\frac{\delta L}
{\delta (2\pi\alpha' A_t')}\right]_{B=0},\label{d7B2}
\end{eqnarray}
where $L$ is the Lagrangian density as defined in (\ref{d7action}). 
This expression follows since the $B$-field appears in the DBI action
(\ref{DBIBaction}) only as a gauge invariant combination of $2\pi\alpha'
F + \hat{B}$. More explicitly they appear as
$(2\pi\alpha'A_t'-\hat{B}_{0r})^2$ in the action, hence we have the
additional negative sign. 
The shape of the D7-branes is specified only by $y(r)$, so
the induced $B$-field is just $\hat{B}_{0r}  = B_{0y} y'(r) + B_{0r}$.
Together with (\ref{constd}), we have 
\begin{eqnarray}
S^{\rm D7}_{\rm DBI}\biggm|_{{\cal O}(B)} = -\bd\int d^4x dr \; \left(
B_{0y} y' + B_{0r}
\right)\, .
\end{eqnarray}
This is the source term for the bulk NSNS B-field.

For our later purpose, we can express $(r,y,z)$ in terms of angular coordinates for the $S^5$ $\{\theta_1,\dots,\theta_5\}$: 
\begin{eqnarray}
z = r_6 \cos\theta_1 \, , \quad
y = r_6 \sin \theta_1 \cos \theta_2 \,, \quad
r = r_6 \sin\theta_1 \sin\theta_2 \, ,\label{Deftheta12}
\end{eqnarray}
the remaining $\theta_3,\theta_4,\theta_5$ parametrize $S^3$ in (\ref{Gab}).
For our D7-brane embedding specified by  $y(r), z=0$, this translates into setting $\theta_1 = \pi/2$. 
We can further invert the relation $r_6^2=r^2+y(r)^2$,
and express $\theta_2$ as a function of $r_6$ via:
\begin{eqnarray}
 \theta_2(r_6) = \arctan \frac{r(r_6)}{y(r(r_6))} \, .
\label{deftheta2r6}
\end{eqnarray} 
So in terms of $r_6$ and these angular 
coordinates, the source coupling for the NSNS $B$-field is 
\begin{eqnarray}
S^{\rm D7}_{\rm DBI}\biggm|_{{\cal O}(B)} 
\!\!\!\! = -\bd\int\! d^4x \!\int \!dr_6 \;d\Omega_5 \;
\frac{\delta(\theta_1-\pi/2)
\delta(\theta_2 - \theta_2(r_6))}{2\pi^2\sin^3\theta_2} 
\left(
B_{0r_6} + B_{0 \theta_2}
\frac{\partial \theta_2}{\partial r_6}
\right).\quad
\label{sourcetheta}
\end{eqnarray}
Note that to incorporate the D7-brane DBI action into the full 10 dimensional supergravity analysis, we have inserted delta-functions which restrict to the specific embedding we are considering. 
In particular we have changed the integral to the 
whole angular coordinates, so we divide it by the volume $V_3=2\pi^2$ of the
unit 3-sphere. 

We will now concentrate on the region $r \sim 0$ to simplify our
situation.  
The spike has a rigid cone shape around the origin $r=0$. 
Around the tip of the cone $r\sim 0$, $y'$ diverges, so 
around there we have 
\begin{eqnarray}
\theta_2\sim \theta_2^{(0)} \equiv 
\frac{\sqrt{\bd^2 - \bc^2}}{\bc}\, .\label{theta20}
\end{eqnarray}
With this, we can approximate the source term (\ref{sourcetheta}) as
\begin{eqnarray}
S^{\rm D7}_{\rm DBI}\biggm|_{{\cal O}(B)}
= -\frac{\bd}{2\pi^2}\int\! d^4x \int \!dr_6 \;d\Omega_5
\;
\delta(\theta_1-\pi/2)
\delta(\theta_2 - \theta_2^{(0)})
\frac{B_{0r_6}}{\sin^3\theta_2^{(0)} }\, .
\label{source}
\end{eqnarray}

\subsubsection{Extracting the RR 3-Form Flux}

We now would like to extract the linearized perturbation $F_3$, which
will be crucial for generating the potential on the instanton moduli
space.   
For this, at the linear order perturbation, it is sufficient to consider
the equations of motion for the NSNS $B$-field, with this
limiting source term (\ref{source}) included,
as other equations are affected only at higher orders (we will
check this later, see eq.~(\ref{f5kin}) and discussions thereafter). 
It will be important that near the $r\sim 0$ region, this source is 
only for $B_{0r_6}$.

The relevant part of the type IIB supergravity action is \cite{Polchinski2}
\begin{eqnarray}
 S_{\rm B} &=& -\frac{1}{4\kappa_{10}^2} \int d^{10}x 
\sqrt{-g_{10}}\; e^{-2\Phi} 
|H_3|^2
+\frac{1}{4\kappa_{10}^2} \int F_5 \wedge B_2 \wedge F_3
\nonumber 
\\
&  &
-\frac{1}{4\kappa_{10}^2} \int d^{10}x\sqrt{-g_{10}}\;\frac12
|\tilde{F_5}|^2 
.
\label{IIBaction}
\end{eqnarray}
Here 
\begin{eqnarray}
 \tilde{F}_5 \equiv F_5-\frac12 C_2\wedge H_3 + \frac12 B_2 \wedge F_3
\, ,
\end{eqnarray}
so, in the third term in (\ref{IIBaction}),  the term linear in $B_2$ is
\begin{eqnarray}
-\frac{1}{8\kappa_{10}^2} \int 
(-C_2\wedge H_3 + B_2 \wedge F_3) \wedge * F_5 \, .
\end{eqnarray}
For a self-dual background 5-form flux $F_5 = * F_5$ (\ref{F5}), this is equal to the second
term of (\ref{IIBaction}).

Substituting the $AdS_5\times S^5$
background metric and the RR 5-form flux (\ref{F5}) 
and writing out in explicit components, we obtain 
\begin{eqnarray}
S_{\rm B} &=& -\frac{1}{2(2\pi)^7 \alpha'^4 g_s^2}
\int d^4x dr_6 d\Omega_5 \;
r_6^3 \left[
H_{0r_6\theta_1}^2 + \frac{1}{\sin^2\theta_2}H_{0 r_6 \theta_2}^2
\right]
\nonumber \\
&&+\frac{1}{(2\pi)^7\alpha'^4} \int d^4x dr_6 d\Omega_5 \; B_{0r_6}F_{123}^{(3)} 2^4 \pi N_c (\alpha')^2\, .
\end{eqnarray}
Here we used the explicit 5-form flux on the $S^5$,  
$F_5 = 2^4 \pi N_c \alpha'^2 d\Omega_5$ (\ref{F5}).
Together with the source action (\ref{source}), the total 
equation of motion for the NSNS $B$-field is
\begin{eqnarray}
0 &= & \frac{r_6^3}{(2\pi)^7\alpha'^4 g_s^2} 
\left[
\p_{\theta_1} \left(
\sin^4 \theta_1 \sin^3 \theta_2 H_{0r_6\theta_1}
\right)
+
\p_{\theta_2} \left(
\sin^2 \theta_1 \sin^3 \theta_2 H_{0r_6\theta_2}
\right)
\right]
\nonumber \\
& & + 
\frac{1}{(2\pi)^7 \alpha'^4}
F_{123}^{(3)} \sin^4\theta_1 \sin^3\theta_2 \; 2^4 \pi N_c \alpha'^2
\nonumber \\
& & - \delta(\theta_1 - \pi/2)
\delta (\theta_2-\theta_2^{(0)})
\frac{\bd}{2\pi^2}\,. 
\label{nsnseq}
\end{eqnarray}
We can recognize this as a 
1+2-dimensional electromagnetism on a compact space spanned by
$\theta_1$ and $\theta_2$. The first term is a total divergence, so the
remaining terms should vanish when we perform an integration over the
2-dimensional space. This condition results in
\begin{eqnarray}
\frac{1}{(2\pi)^7 \alpha'^4}
F_{123}^{(3)} 
\int_0^\pi\! d\theta_1 \; \sin^4\theta_1 
\int_0^\pi\! d\theta_2 \; \sin^3\theta_2 \; (2^4 \pi N_c \alpha'^2)
= \frac{\bd}{ 2\pi^2}\, .
\end{eqnarray}
Performing the integration and re-arranging, 
we obtain the constant RR 3-form flux $F_{123}^{(3)}$ as given in (\ref{f3back}).
This is the leading order effect of the backreaction of the D7-brane electric
flux. Note that the supergravity equation of motion for the $F_3$ flux 
is trivially satisfied with this constant configuration.

Interestingly, this result (\ref{f3back})
is the same one obtained previously by
solving the $F_3$ equation of motion with the smeared baryon vertices in 
Sec.~\ref{sec3.1}. Here we have not
assumed the presence of the baryon vertices, 
but the supergravity equation of
motion ``knows'' the presence for its consistency.

\section{Dissolution of the Instanton and Color-Flavor Locking}
\label{sec4}

In this section, we show that the backreacted supergravity flux
(\ref{f3back}) generates a nontrivial potential for the instanton moduli
space (Sec.~\ref{sec41}). 
It provides a dynamical mechanism for fattening the size of the
instanton on the flavor D7-branes. We compute the potential explicitly
in Sec.~\ref{sec42}.

This is a dynamical color-flavor locking in the holographic QCD, because
the size of the instanton on the D7-branes 
is the vev of the squark of the supersymmetric QCD. 
Since the squarks
are in the bi-fundamental representation of the color and the flavor
symmetries, their condensation gives a color-flavor locking. 
The squark condensation means that the
theory favors Higgs phase when the baryon 
chemical potential $\mu$ is larger
than the quark/squark mass $m$. 
We study the patterns of the symmetry breaking in Sec.~\ref{sec43}.

\subsection{Additional Chern-Simons Term}
\label{sec41}
\paragraph{}
Using the backreacted supergravity solution (\ref{f3back}), 
we obtain an additional Chern-Simons term induced on the
D7-branes. The general formula for the 
Chern-Simons couplings on the D7-branes is  
\begin{eqnarray}
 \mu_7 \;{\rm tr} \int \exp(2\pi\alpha' F + B_2)\wedge \sum_q C_q
\, .\label{CSD7}
\end{eqnarray}
Here the D7-brane RR charge is $\mu_7  = \frac{1}{(2\pi)^7 \alpha'^4}$, and the field strength $2\pi\alpha' F$ now also contains non-Abelian instanton piece. 
Formally expanding (\ref{CSD7}) out and performing integration by parts, 
for non-zero $F_{123}^{(3)}$ we obtain the explicit expression in components
\begin{eqnarray}
S_{\rm CS} = \frac{1}{8 (2\pi)^4 \alpha'}
\int d^4 x F_{123}^{(3)}
\int d^4\xi \;
{\rm tr}
\left[A_0 F_{ij} F_{kl}\epsilon^{ijkl}\right]+\cdots\label{CS}
\end{eqnarray}
Here $\cdots$ means terms necessary to form a gauge-invariant CS 5-form
\begin{eqnarray}
 {\rm tr}\left[
AFF -\frac{1}{2} A^3 F + \frac{1}{10} A^5
\right]
\end{eqnarray}
where the wedge product $\wedge$ is omitted.
The second and the third terms in the CS action are
irrelevant for our subsequent discussions. We substitute the
constant $F_{123}^{(3)}$ from the linearized supergravity backreaction
(\ref{f3back}) to extract the relevant term from $S_{\rm CS}$:
\begin{eqnarray} 
\frac{\alpha'\bd}{16\pi N_c}
\int d^4 x \int d^4\xi \; 
{\rm tr}
\left[A_0 F_{ij} F_{kl}\epsilon^{ijkl}\right].\label{ACSterm}
\end{eqnarray}
This is the leading correction term due to the baryon density ${\bf d}$. Note
that the factor $1/N_c$ in front of above shows that this is
indeed a correction to the original D7-brane action. %{\bf (Check this!)}
%
%\if0
\begin{figure}[t]
\begin{center}
\includegraphics[width=0.7\textwidth]{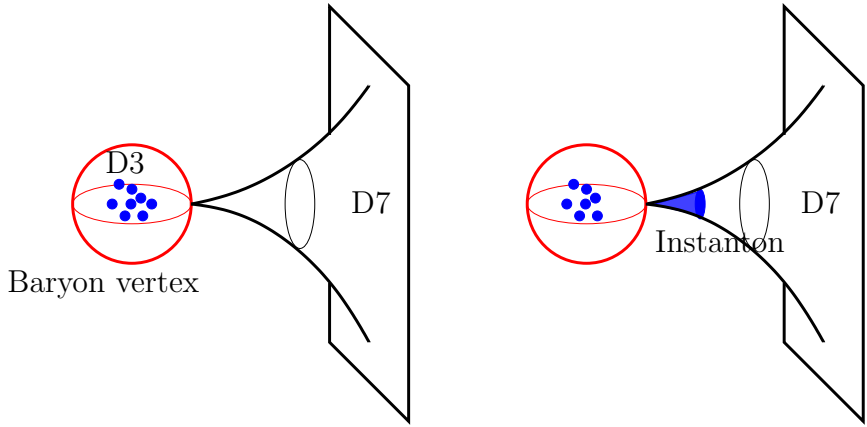}
\put(-195,80){D7}
\put(-25,80){D7}
\put(-325,50){Baryon vertex}
\put(-288,95){D3}
\put(-80,65){Instanton}
\caption{ 
A schematic picture of the charge conservation process.
In the left figure, $N_c$ D3-branes are sitting inside the baryon vertex
(D5-brane wrapping $S^5$). From the D3-branes, $N_c$ units of RR 5-form
flux emanates. The CS term on the D5-brane creates an electric charge
on the worldvolume of the D5-brane \cite{WittenBaryon}, and this generates
electric field on the spiky D7-brane which touches the D5-brane.
In the right figure, we move one D3-brane toward outside of the baryon
vertex. This D3-brane becomes an instanton (shaded region on the
 D7-brane spike). The instanton is electrically charged, so the total
 electric flux going to the asymptotic infinity of the D7-brane
 worldvolume is conserved.
}
\label{figcharge}
\end{center}
\end{figure}
%\fi
%

This additional CS term has an important 
physical meaning. The essence here is quite similar to the
generation of the baryon charge in the Sakai-Sugimoto model
\cite{SaSu1}, while the use of the instantons here is quite different
from there ({\it c. f.} footnote 6.).
As discussed in Sec.~\ref{sec1}, we are studying the process
of moving one D3-brane from the origin onto the worldvolume of the
D7-branes. Once the single D3-brane goes outside the D5-brane wrapping
the $S^5$, the RR 5-form flux penetrating the $S^5$ worldvolume of the
D5-branes reduces by one unit. This 5-form was responsible for the CS
term on the D5-branes to produce the electric charges on the D5-branes,
which are the end points of the fundamental strings. So, by this moving
process, the total number of the fundamental strings decreases by a
fraction of $1/N_c$. Then, where does the baryon charge go off to?
The answer is the new CS term (\ref{CS}). Once the D3-brane gets on the
D7-branes, it induces an instanton charge.
The instanton number for the single instanton is 
\begin{eqnarray}
{\rm tr} \int d^4\xi \;
F_{ij} F_{kl}\epsilon^{ijkl} = 32\pi^2.
\end{eqnarray}
For a small size instanton, the CS term (\ref{CS}) effectively becomes proportional to 
\begin{eqnarray}
\frac{2\pi\alpha' \bd}{N_c}\int d^4x d^4\xi \; {\rm tr}A_t \; \delta^4(\xi) \, .
\end{eqnarray}
This means that the instanton (which is the D3-brane dissolved into the
D7-brane) carries the electric charge $(2\pi\alpha' \bd)/N_c$. Compared to this amount of the charge, the
original solution (\ref{KOsol}) 
shows that the spike has an electric charge $2\pi \alpha' \bd$.
We can therefore conclude that,
pulling out one from the
$N_c$ D3-branes decreases the baryon charge by its fraction $1/N_c$. 
This decrease is indeed offset by the instanton sourcing the electric
field, so that the asymptotic expression for the electric field doesn't
change. {See Fig.~\ref{figcharge} for a schematic explanation of this
charge conservation.}

\subsection{Potential on the Instanton Moduli Space}
\label{sec42}

Finally we have collected all the pieces for computing the induced potential on the instanton moduli space,
and we will show that it drives the instanton(s) into dissolution. 
Here we consider a simpler special case, where 
the BPST instanton profile is centered at the origin: 
\begin{eqnarray}
 {\rm tr}
F_{ij} F_{kl}\epsilon^{ijkl} = \frac{192 \rho^4}{(\tilde{r}^2 +
  \rho^2)^4}\,. 
\end{eqnarray}
Note that the BPST instanton solution is obtained in the conformally flat $\tilde{r}$ coordinate, not in the $r$ coordinate.
Upon substituting into the additional CS term (\ref{ACSterm}), we can
extract the potential for the size $\rho$, $V_B(\rho)$, via the relation
$S=-\int d^4x V(\rho)$ as
\begin{eqnarray}
V_B(\rho) = -\frac{12 \rho^4 (2\pi\alpha'\bd)}{N_c} \int_{\frac{r_0}{2^{1/3}}}^\infty
\!\! d\tilde{r} \; A_t(r) \frac{\tilde{r}^3}{(\tilde{r}^2 + \rho^2)^4}
\, .
\end{eqnarray}
Again, note that the argument of the electric potential $A_t$ is $r$
which is related to $\tilde{r}$ by (\ref{rtrrelation1}).
Integrating by parts (for the $\tilde{r}$ coordinate) and use (\ref{KOsol}), we obtain
\begin{eqnarray}
 V_B(\rho) &=&
-\frac{2\pi\alpha'\bd}{N_c}
\int_0^\infty \! dr \;
A'_t(r)
\frac{\rho^4 (3\tilde{r}^2 + \rho^2)}{(\tilde{r}^2 + \rho^2)^3} 
\nonumber \\
& = &-\frac{2\pi\alpha'\bd}{{\cal N}N_c}
\int_0^\infty \! dr \;
\frac{\bd}{\sqrt{ r^6+r_0^6}}
\frac{\rho^4 (3\tilde{r}^2(r) + \rho^2)}{(\tilde{r}^2(r) + \rho^2)^3} 
\, .
\label{potrho}
\end{eqnarray}
%In the last equation, we substituted the $U(1)$ solution (\ref{KOsol}). 
This (\ref{potrho})
is indeed a monotonically decreasing function of $\rho$, when viewing 
together with the coordinate redefinition (\ref{rtrrelation1}).
This can be easily understood if we notice following three facts: (i) $A_t'(r)$ is a
monotonically decreasing function of $r$. (ii) The last factor in the
integrand of (\ref{potrho}) is the instanton density function which is
peaked at $\tilde{r}=0$ and monotonically decreasing in $\tilde{r}$,
while the width of the function is given by $\rho$ and the function has
a normalized integral (which is the instanton number). (iii) The map
(\ref{rtrrelation1}) between $r$ and $\tilde{r}$ is a one-to-one and
monotonic function.
The instanton modulus potential $V(\rho)$ (\ref{potrho}) shows that the system
dynamically favors the Higgs phase, $\rho\neq 0$. This is the
color-flavor locking in the supersymmetric QCD. 

The $V_B(\rho)$ from the CS term computation (\ref{potrho}) does not appear to be analytically integrable,  however to get a qualitative understanding we can consider the following asymptotic values:
\begin{eqnarray}
 V_B(\rho=0) = 0\, , \quad
 V_B(\rho=\infty) = -\frac{2\pi\alpha'\bd}{N_c} 
\int_0^\infty A_t' dr = -\frac{2\pi\alpha'{\bf d}\mu}{N_c}\, .
\end{eqnarray}
Their difference, 
\begin{eqnarray}
 V_B(\rho=0)-V_B(\rho=\infty) = \frac{2\pi\alpha'\bd\mu}{N_c}
\label{height}
\end{eqnarray}
is consistent with the interpretation that
we pull out one quark
per each baryon to the infinity in the background
chemical potential $\mu$.
However, note that the constant $F_{123}^{(3)}$ is obtained only in the
vicinity of $r=0$. So, our calculation is strictly valid only for small $\rho$, and
the potential height (\ref{height}) derived at large $\rho$ should not be reliable. We however expect that the qualitative physical result
(\ref{height}) is not modified significantly when we taken into account of full $F_{123}^{(3)}$ at large radius.
For completeness, here we include the plot of the one instanton size modulus potential, normalized by the asymptotic value $V_B(\infty)$:
%
%\if0
\begin{figure}[ht]
\begin{center}
\includegraphics[width=0.45\textwidth]{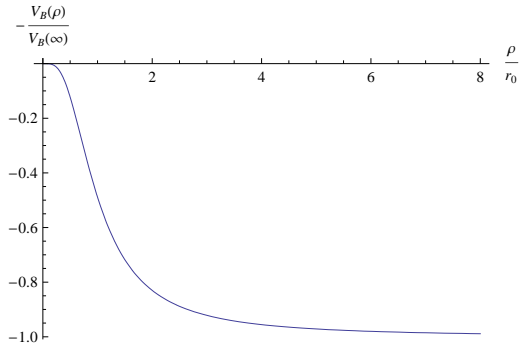}
\caption{The plot of $\frac{V_B(\rho)}{|V_B(\infty)|}$ versus
 $\frac{\rho}{r_0}$} 
\label{figrunawayT=0}
\end{center}
\end{figure}
%\fi
%%%%%%%%%%%%%%%%%%%%%%%%%%%%%%%%%%%%%%%%%%%%%%%%%%%%%%%%%%%%%%%%%%%%%%%%%%%%%%%% 
In the analysis above, we assumed that the center of the BPST instanton
is at the origin. However, we can consider a generic situation where the
center of the instanton is not at the origin $r=0$. Suppose that the
center is at some distance ${\bf X}$ from the origin. It is
clear that the similar expression to (\ref{potrho}), which is again only valid at small $r$, would give a qualitative result that the
potential $V_B(\rho,{\bf X})$ goes to its minimum at 
$(\rho,{\bf X})=(\infty, {\rm finite})$ or 
$(\rho,{\bf X})=({\rm finite}, \infty)$. The latter is in particular an
extreme point in the moduli space in the Coulomb phase. There $\rho$ can
vanish and in that case the color-flavor locking does not occur. However,
for ${\bf X}\neq 0$, the original gauge group $U(N_c)$ is explicitly
broken, as it is in the Coulomb phase.

Instead of substituting the BPST instanton, we can substitute
multi-instanton solutions. Suppose we treat $\tilde{N}_c$ instantons.
We need to require $\tilde{N}_c \ll N_c$, because in the
perturbation of the backreaction the original background of
$AdS_5\times S^5$ should not be drastically modified.
Generically, the instantons are separated from each
other.\footnote{Since electrically charged instantons should repel each
other, generic configurations should be with the separated instantons.
This phenomenon is common with the interaction among baryons 
\cite{Hashimoto:2009ys} in the Sakai-Sugimoto model \cite{SaSu1}; the
repulsive core of nucleons is mainly  
due to this electric repulsion.}
Our analysis for the single BPST instanton holds also for the
multi-instanton case. Then it is shown that all the size moduli of the
instantons are destabilized. 

\subsection{What is Locked?}
\label{sec43}

Naively speaking, the condensation of the squark field which comes from 
a fluctuation of a string connecting the D3-branes and the D7-branes 
gives the
instanton size. This string transforms in the fundamental representation
of the gauge group $U(\tilde{N}_c)$ and in the anti-fundamental
representation of the 
flavor group\footnote{
Precisely speaking, the flavor group of the supersymmetric QCD 
should be $SU(N_f)$, since the overall $U(1)$ transformation of the
flavor symmetry can be identified as a global part of a $U(1)$ subgroup
of the local gauge group. In the following, we adopt $U(N_f)$ rather
than $SU(N_f)$, but the argument goes similarly.
} $U(N_f)$. Here $U(\tilde{N}_c)$ is a part of the total
gauge group $U(N_c)$ {\it i.e.} $\tilde{N}_c \leq N_c$. The partial
color symmetry is for our convenience to consider $\tilde{N}_c$
instantons only to dissolve into the worldvolume of the D7-branes.
By the squark condensation, apparently 
the color and the flavor symmetries
$U(\tilde{N}_c)\times U(N_f)$ 
are broken. We shall see the symmetry breaking pattern.

The theory on the instantonic D3-branes is constructed from their ADHM
data \cite{Douglas:1995bn}. 
The ADHM data consists of four $U(\tilde{N}_c)$ adjoint scalar
fields which are combined into two 
$\tilde{N}_c\times \tilde{N}_c$ complex
scalar fields $B_1$ and $B_2$, and the squark fields $I^\dagger$ and $J$
which are complex $N_f\times \tilde{N}_c$ matrix scalar fields. The
squark fields transform as 
\begin{eqnarray}
 I^\dagger \mapsto U I^\dagger U_0^{-1}, 
\quad
 J \mapsto U J U_0^{-1}, 
\end{eqnarray}
where $U\in U(N_f)$ and $U_0 \in U(\tilde{N}_c)$. So these fields are
in the bi-fundamental representation. %, and identified as a part of the 
%squark fields in QCD. 

Let us consider 't Hooft instantons. Then $B_1$ and $B_2$ are
diagonal matrices, and the ADHM
equations are nothing but the BPS equations for the theory on the
D3-branes, are
\begin{eqnarray}
I I^\dagger = J^\dagger J \, , \quad I J =0 \, .
\end{eqnarray}
This of course allows a trivial solution $I^\dagger = J =0$, which
corresponds to the zero-size instanton. 

First 
we consider the two-flavor case $N_f=2$. The flavor
symmetry is $U(2)$ which is a vector part of the chiral
symmetry (the chiral symmetry is explicitly broken by the quark/squark 
mass in our case). The 'tHooft instanton 
whose centers are located at the origin is 
represented by a solution 
\begin{eqnarray}
 I^\dagger = 
\left(
\begin{array}{cccc}
 \rho_1 & \rho_2 & \cdots & \rho_{\tilde{N}_c}\\
0 & 0 & \cdots & 0 
\end{array}
\right), \quad 
J = 
\left(
\begin{array}{cccc}
0 & 0 & \cdots & 0 \\
\rho_1 & \rho_2 & \cdots & \rho_{\tilde{N}_c}
\end{array}
\right).
\label{ij}
\end{eqnarray}
Here, all $\rho_i$'s are real parameters.
These correspond to size of each instanton. 
With this at hand, we can compute unbroken symmetry.

For simplicity, we 
consider the case of a single instanton, $\tilde{N}_c=1$.
\begin{eqnarray}
 I^\dagger = 
\rho \left(
\begin{array}{cccc}
 1\\
0 
\end{array}
\right), \quad 
J = \rho
\left(
\begin{array}{cccc}
0 \\
1
\end{array}
\right).
\end{eqnarray}
Here $\rho$ is a nonzero constant (for which we computed the potential).
In this case, the transformation which leaves $I^\dagger$ intact is
\begin{eqnarray}
U = \left(
\begin{array}{cc}
e^{i\alpha_1} & 0 \\
0 & e^{i\alpha_2}
\end{array}
\right) , \quad U_0 = e^{i\alpha_1}, \quad \alpha_i \in {\mathbb R}
\, .
\end{eqnarray}
On the other hand, the symmetry which leaves $J$ intact is
\begin{eqnarray}
U = \left(
\begin{array}{cc}
e^{i\alpha_1} & 0 \\
0 & e^{i\alpha_2}
\end{array}
\right) , \quad U_0 = e^{i\alpha_2},
\quad \alpha_i \in {\mathbb R}\, .
\end{eqnarray}
Therefore, we need to require $\alpha_1=\alpha_2$, and we conclude,
\begin{eqnarray}
U(1)_{\rm color} \times  U(2)_{\rm flavor} \to U(1)_{\rm CFL} \, .
\end{eqnarray}
The global part of $U(1)_{\rm color}$ is locked with the diagonal part
of $U(2)_{\rm flavor}$, which is the baryonic symmetry $U(1)_B$, and  
the local part of $U(1)_{\rm color}$ is broken.

This kind of the color-flavor locking can be found for general $N_f$. 
In the case of $\tilde{N}_c=1$, the squark condensations are given by
\begin{eqnarray}
 I^\dagger = 
\rho \left(
\begin{array}{cccc}
 1\\
0 \\
0\\
\vdots \\
0
\end{array}
\right), \quad 
J = \rho
\left(
\begin{array}{cccc}
0 \\
1 \\
0 \\
\vdots\\
0
\end{array}
\right).
\end{eqnarray}
A similar analysis shows
\begin{eqnarray}
U(1)_{\rm color}  \times  U(N_f)_{\rm flavor}
\to U(1)_{\rm CFL} \times U(N_f-2)_{\rm flavor}\, .
\end{eqnarray}
$U(1)_{\rm CFL}$ is a global symmetry which locks a part of the
flavor symmetry and the gauge symmetry. Note that for $N_f>2$, this
$U(1)_{\rm CFL}$ can be nothing to do with 
the baryonic symmetry $U(1)_{B}$, since
the action of this 
$U(1)_{\rm CFL}$ can be chosen as 
\begin{eqnarray}
 U = \mbox{diag}(e^{i\alpha}, e^{i\alpha},
e^{-2i\alpha/N_f},
\cdots,
e^{-2i\alpha/N_f}), 
\quad U_0 = e^{i\alpha}, \quad \alpha\in {\mathbb R} \, .
\end{eqnarray}

For generic $\tilde{N}_c$, 
we expect that all the size moduli are driven to have nonzero values
(for example, for $N_f=2$ we have (\ref{ij})).
So the symmetry is broken as
\begin{eqnarray}
U(\tilde{N}_c)_{\rm color} \times   U(N_f)_{\rm flavor}
\to U(1)_{\rm CFL} \times U(N_f-2)_{\rm flavor}\, .
\end{eqnarray}
The locking is quite similar to the case of $\tilde{N}_c=1$. 
Note that we restrict ourselves to the case of the 'tHooft instantons.
Since the 
'tHooft instantons do not cover all the moduli space of the instantons,
there remains 
a possibility that the unbroken symmetry, in particular the
part concerning the gauge symmetry, may be enhanced.

In the D-brane analysis, we used the technique for Coulomb branch in
AdS/CFT correspondence and treat one D3-brane as a probe by separating
it from the rest, by hand. This procedure for the separation is 
somewhat artificial, but it is required for the geometry not to be
drastically modified by a possible back reaction, which is our
limitation.

\section{Extension to Finite Temperature System}
\label{sec5}

In this section we explore the system at finite temperature and baryon density.
Since the boundary geometry of our $AdS_5\times S^5$ is $R^{1,3}$, the geometry
creates a horizon inside $AdS$ at any finite temperature.
$\cN=4$ SYM theory coupled with $\cN=2$ hypermultiplets in fundamental representation at finite temperature
has been studied (See \cite{Erdmenger:2007cm} for a review).
As mentioned in Sec.~\ref{sec1}, there are two brane embeddings in the black hole background:
the Minkowski embedding and the black hole embedding (Fig.~\ref{figembed}).
The former describes 
the configuration of D-branes staying outside of the horizon everywhere
and the latter describes 
the configuration of D-branes falling into the black hole horizon. 
The equations of motion and the free energy determine which configuration is realized at
given quark mass and baryon density normalized by temperature: $m/T$ and $2\pi\alpha'{\bf d}/T^3$.

In the case of zero baryon density and no instanton, 
the Minkowski embedding covers only higher $m/T$ region while
the black hole embedding covers only lower $m/T$ region \cite{Mateos:2006nu}.
These two are connected by a first order phase transition at certain critical temperature $(m/T)_{\rm crit.}$.
This phase transition is interpreted in the field theory as meson melting:
the spectrum is discrete and the mesons are stable in the Minkowski
embedding, 
while the spectrum is continuous and the mesons are unstable in the black hole embedding.
In the case of finite baryon density, the Minkowski embedding is no longer physically allowed 
and the black hole embedding covers the whole temperature region.
\if0
{\bf This canonical ensemble picture does not cover the whole phase space.
More complete picture is obtained by considering the grand canonical ensemble
 \cite{HoloBaryon,Karch:2007br,D3D7phase,Nakamura:2006xk,Erdmenger:2007ja}.
}
\fi
When an instanton is excited on the D7-branes at zero baryon density,
the potential for the instanton size moduli takes its minimum at the origin, $\rho=0$,
for the Minkowski embeddings and at some finite value, $\rho=\rho_{\rm min}>0$,
for the black hole embeddings \cite{Apreda1}.
This means the system is in Higgs phase above the critical temperature $(m/T)_{\rm crit.}$.
{Therefore, the system is already in a CFL phase with the squark
condensation, in the melted meson phase with finite $T$.}

The purpose of this section is to study the case with both the finite
baryon density and the instanton configuration.
As we saw in the zero temperature case, the backreaction of the baryon density excites an additional CS term
which induces the CFL.
We will see how this CS term affects the instanton potential in the
finite temperature system.

\subsection{The D3D7 System at Finite Baryon Density and Temperature}
\label{sec51}

The background geometry dual to the finite temperature system is an AdS black hole.
In Poincare like coordinates, the metric, RR 4-form and the dilaton have 
the following expressions, in the conventions of Ref.~\cite{HoloBaryon}:
\begin{eqnarray}
ds^2&=&{1\over2}{u^2\over R^2}
\left(-{f^2\over \wt{f}}dt^2+\wt{f} dx_3^2\right) 
+{R^2\over u^2}\left(du^2+u^2 ds_5^2\right)  \cr
C&=&{1\over R^4}\left( {u^2\over2}+ {u^4_0\over 2 u^2} \right)^2 d^4x\, 
,~~~~~~e^{\Phi}=e^{\Phi_0}\, ,
\end{eqnarray}
where
\begin{eqnarray}
f=1-{u_0^4\over u^4}\,,~~~\wt{f}=1+{u_0^4\over u^4}\,,
\end{eqnarray}
with $u_0$ the location of the horizon. 
$u$ and $r_6$ in Sec.~3 are related to each other by the coordinate
transformation  
\begin{eqnarray}
u^2=r_6^2+\sqrt{r_6^4-u^4_0} \, .
\end{eqnarray}
The regularity of the Euclidean section of this geometry relates the horizon radius $u_0$
and the Hawking temperature, which is interpreted as a temperature of
the boundary gauge theory of our concern, as
\begin{eqnarray}
T={u_{0}\over \pi R^2}\, .
\end{eqnarray}
As in the zero temperature case, we foliate $S^5$ with $S^3$ so that $SO(3)$ R-symmetry is manifest:
\begin{eqnarray}
du^2+u^2ds_5^2
&=&du^2+u^2(d\theta^2+\sin^2{\theta}ds^2_3+\cos^2\theta d\phi^2)\, ,
\end{eqnarray}
where
$\theta,\phi$  and $\theta_1, \theta_2$ in (\ref{Deftheta12}) are
related through the following equations
\begin{eqnarray}
\sin\theta=\sin\theta_1\sin\theta_2\, ,
~~~~~\tan\phi={\cos\theta_1\over \sin\theta_1\cos\theta_2}\, .
\end{eqnarray}
The probe D7-brane worldvolume is spanned by $(t,x^i,u,S^3)$, and
is localized in $\phi$ direction, which we can use rotational symmetry
to set $\phi=0$ (corresponding to $\theta_1=\frac{\pi}{2}$). 
In the new coordinates, the embedding is described by $\chi\equiv \cos\theta$ as a function of $u$.
The $U_b(1)$ gauge field $A$ dual to the baryon current on the boundary has only non-zero time component
\begin{eqnarray}
A=A_t(u)dt.
\end{eqnarray} 
With these ans\"atze, the DBI action of the D7-branes is given by
\begin{eqnarray}
 \frac{S_{\rm DBI}^{\rm D7}}{V_4}
&=&\int\! du \; \cL\cr
 &=&  -\cN \int\! du \; {u^3f\tilde{f}(1-\chi^2)\over4}
 \sqrt{1-\chi^2+u^2\dot{\chi}^2-(2\pi\ap \dot{A_t})^2{2\tilde{f}(1-\chi^2)\over f^2}} 
\end{eqnarray}
where the dot denotes ${d\over du}$ and ${\cal N}$ is as defined below
(\ref{d7action}). 
Since the action does not contain $A$ explicitly, the momentum conjugate of $A$ is constant:
\begin{eqnarray}
{\delta \cL\over \delta(2\pi\ap  \dot{A}_t)}=\cN {u^3\over 2}{\tilde{f}^2\over f}
{(1-\chi^2)^2 (2\pi\ap\dot{A}_t)\over  \sqrt{1-\chi^2+u^2\dot{\chi}^2-(2\pi\ap \dot{A_t})^2{2\tilde{f}(1-\chi^2)\over f^2}} }
  \equiv {\bf D}\, , 
\end{eqnarray}
or equivalently,
\begin{eqnarray}
2\pi\ap\dot{A}=
2\left({{\bf D}\over \cN}\right)
{f\sqrt{1-\chi^2+u^2\dot{\chi}^2} \over \sqrt{\tilde{f}(1-\chi^2)}
\sqrt{u^6\tilde{f}^3(1-\chi^2)^3+8({\bf D}/\cN)^2 }}\, .
\label{ei-dotto}
\end{eqnarray}
To derive the equation of motion for $\chi$, we Legendre transform the action with respect to ${\bf D}$
so that $\dot{A}_t$ can be eliminated from the action:
\begin{eqnarray}
\tilde{\cL}&=&\cL-{\delta L\over \delta (2\pi\ap \dot{A}_t)}(2\pi\ap \dot{A}_t)\cr
&=&-{\cN\over 4}{f\over \sqrt{\tilde{f}}\sqrt{1-\chi^2}}
\sqrt{1-\chi^2+u^2\dot{\chi}^2}
\sqrt{u^6\tilde{f}^3(1-\chi^2)^3+8({\bf D}/\cN)^2}\, .
\end{eqnarray}
Then the $\chi$ equation is
\begin{eqnarray}
&&\pd_{u}
\left(
{u^5f\tilde{f}(1-\chi^2)\dot{\chi}\over
\sqrt{1-\chi^2+u^2\dot{\chi}^2}}\sqrt{1+{8({\bf D}/\cN)^2\over
u^6\tilde{f}^3(1-\chi^2)^3}} 
\right)\cr 
&& \quad 
=-{u^3f\tilde{f}\chi\over \sqrt{1-\chi^2+u^2\dot{\chi}^2}}
\sqrt{1+{8({\bf D}/\cN)^2\over u^6\tilde{f}^3(1-\chi^2)^3}}\cr
&&\quad\quad 
\times\left(
3(1-\chi^2)+2u^2\dot{\chi}^2-24\left({{\bf D}\over \cN}\right)^2{1-\chi^2+u^2\dot{\chi}^2 \over 
u^6\tilde{f}^3(1-\chi^2)^3+8({\bf D/\cN})^2}
\right).
\end{eqnarray}
%%%%%%%%%%%%%%%%%%%%%%%%%%%%%%%%%%%%%%%%%%%%%%%%%%%%%%%%%%%%%%%%%%%%%%%%%%%%%%
As studied in Ref.~\cite{Frolov:2006tc}, the boundary condition of probe branes at the horizon is determined by the regularity of the induced curvature:
$\dot{\chi}|_{u=u_0}=0$. With this boundary condition,
the solution of this equation of motion near the horizon $u\sim u_0$ is then given by
\begin{eqnarray}
\chi=\chi_0 -{3\chi_0(1-\chi_0^2)^3\over 4(({\bf D}^2/\cN^2u_0^3)+1-\chi^6_0-3\chi_0^2(1-\chi^2_0))}\left(\!{u\over u_0}\!-\!1\!\right)^2\!
+\cO\left(\!\left({u\over u_0}\!-\!1\right)^3\right)\, .
\end{eqnarray}
Therefore, the embedding can be approximated as
\begin{eqnarray}
\chi=\chi_0,~~~\dot{\chi}=0,
\label{flat-approx}
\end{eqnarray}
for ${u\over u_0}-1$ smaller than ${2\over \sqrt{3}(1-\chi_0^2)^{3/2}}
{{\bf D}\over \cN u_0^{3/2}}$ with large
${\bf D}$.
%%%%%%%%%%%%%%%%%%%%%%%%%%%%%%%%%%%%%%%%%%%%%%%%%%%%%%%%%%%%%%%%%%%%%%%

We again consider the instanton excitations as a perturbation in this background field.
At the leading order of ${\bf D}/N_c$, the instanton couples to RR
4-form and the induced metric, and the relevant terms are: 
\begin{eqnarray}
S_{\rm DBI}^{\rm D7}(FF)&=&-N_f{\cal T}_{\rm D7}\int d^4x \int {u^4\over
4R^4}f\wt{f} \cdot{(2\pi \alpha')^2\over 8}\Tr[F\wedge F] \, ,
\label{therm-poten-DBI}
\end{eqnarray}
\begin{eqnarray}
S_{\rm CS}^{\rm D7}(FF)=N_f{\cal T}_{\rm D7}\int d^4x\int {u^4\over
4R^4}\tilde{f}^2\cdot 
{(2\pi \alpha')^2\over 8} \Tr[F\wedge F] \,. 
\label{therm-poten-CS}
\end{eqnarray}
The instanton $F\wedge F$ lives on an effective four dimensional space whose metric is
\begin{eqnarray}
\wt{G}^4_{ij}&=&\left({1\over2}\sqrt{f\wt{f}}\right)
\left(\left({1-\chi^2+u^2 \dot{\chi}^2 \over 1-\chi^2}+{(2\pi
\alpha')^2\dot{A}^2\over -{1\over2}{f^2\over \wt{f}}}\right)du^2 
+u^2(1-\chi^2)ds_3^2\right) \cr
&=&{\sqrt{f\wt{f}} \over2}
\left(
\left(1+{u^2 \dot{\chi}^2 \over 1-\chi^2}\right)
{u^6\tilde{f}^3(1-\chi^2)^3 \over u^6\tilde{f}^3(1-\chi^2)^3+8({\bf
D}/\cN)^2}du^2 
+u^2(1-\chi^2)ds_3^2\right)\,.\nonumber\\ 
\label{eff-4-met}
\end{eqnarray}
Note that ${\bf D}$ dependence appears only through this metric.
The difference between the DBI term and the CS term is the factors of $f$ and $\tilde{f}$
in the integrands. Therefore, the instanton potential vanishes as long as the temperature is zero,
even in the presence of finite baryon density as we have seen.

\subsection{CS Term from Backreaction}\label{CSback}
\label{sec52}

The next leading order in ${\bf D}/N_c$ comes from the backreaction of
the gauge field to RR fields. 
Similar calculations show that it excites the same constant RR 3-form field near the horizon as that of the zero temperature case with
${\bf d}$ replaced by ${\bf D}$,
\begin{eqnarray}
F^{(3)}_{123}={8\pi^3\ap^2{\bf D}\over N_c}\, .
\end{eqnarray}
Therefore it induces the same CS term 
\begin{eqnarray}
S_{\rm CS}^{\rm D7}({\bf{D}})={\ap{\bf D}\over 16\pi N_c}\int d^4x \int
\tr\left[A\wedge F\wedge F\right] .
\label{backre-poten-CS}
\end{eqnarray}
The remaining issue is to obtain the explicit 
instanton configuration $\tr\left[ F\wedge F\right]$ on the effective
metric $\wt{G}_{ij}^{(4)}$. 
As in the zero temperature case, we would like to obtain a conformally flat coordinate $\wt{u}$ satisfying:
\begin{eqnarray}
\wt{G}_{ij}^{(4)}=S(\wt{u})^2(d\wt{u}^2+\wt{u}^2ds_3) \, .
\label{conf-flat-met}
\end{eqnarray}
For the approximate embedding near the horizon (\ref{flat-approx}), the conformally flat metric (\ref{conf-flat-met}) 
is obtained by
\begin{eqnarray}
{u^2(1-\chi_0^2)\sqrt{\tilde{f}}\over \sqrt{u^6\tilde{f}^3(1-\chi^2_0)^3+8({\bf D}/\cN)^2}}du=
{d\tilde{u}\over \tilde{u}}\, .
\end{eqnarray}
With this conformally flat coordinate, the BPST instanton is given by
\begin{eqnarray}
\tr\left[ F\wedge F\right]={192\rho^4\over
(\wt{u}^2+\rho^2)^4}d^4\wt{\xi}\, . 
\end{eqnarray}

As we have seen, there are three terms contributing the instanton potential for the finite temperature case:
the DBI and the CS term for thermal effect, which we denote $V_T$, and the CS term from the backreaction,
which we denoted earlier as $V_B$.
From (\ref{therm-poten-DBI}) and (\ref{therm-poten-CS}),
the thermal potential is 
\begin{eqnarray}
V_T(\rho)&=&-(S_{\rm DBI}^{\rm D7}(FF)+S_{\rm CS}^{\rm D7}(FF))/V_4 \cr
&=&-{\cN \over 4R^4}
\int \! d\wt{u} \; u^4 \wt{f}(\wt{f}-f) \cdot{(2\pi \alpha')^2\over 8}
{192\rho^4 \tilde{u}^3\over (\wt{u}^2+\rho^2)^4} \, .
\label{thermal-pot-fini-T}
\end{eqnarray}
This potential has a minimum at finite $\rho$ since
$V_T(0)=V_T(\infty)=0$. 
On the other hand, from (\ref{backre-poten-CS}), 
the potential from the backreaction is
\begin{eqnarray}
V_B(\rho)&=&-S_{\rm CS}^{\rm D7}({\bf D})/V_4 \cr
&=&-{{\bf D}^2\over N_c\cN }
\int_{u=u_0}du
{ 2f\over \sqrt{\tilde{f}} \sqrt{u^6\tilde{f}^3(1-\chi_0^2)^3+8({\bf
D}/\cN)^2 }}
{\rho^4(3\wt{u}^2+\rho^2)\over(\wt{u}^2+\rho^2)^3}
\, . 
\label{back-reac-pot-fini-T}
\end{eqnarray}

The approximation we used to obtain the integrands breaks down for large
$u$.  
However, the following two features may still hold even beyond the
approximation: 
$V_B(0)=0$ and $V_B(\infty)=-{(2\pi\ap){\bf D}\mu\over N_c}$.
The first one, $V_B(0)=0$, comes from the fact that the integration of the instanton term,
the last term in (\ref{back-reac-pot-fini-T}), is zero for $\rho=0$. Therefore, independent of
the form of $F_3$ and $\dot{A}_t$, the integration gives zero.
The second one, $V_B(\infty)=-{(2\pi\ap){\bf D}\mu\over N_c}$, comes from the physical reason
explained in Section 4.2 for the zero temperature case.
A numerical analysis suggests that $V_B$ monotonically decreases from
zero to $-{(2\pi\ap){\bf D}\mu\over N_c}$.

\begin{figure}[ht]
\begin{center}
\includegraphics[width=0.45\textwidth]{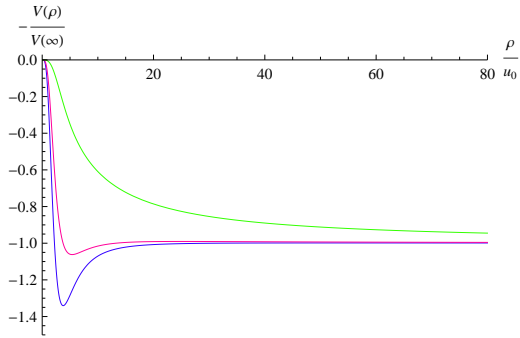}
\caption{The plot of $\frac{V(\rho)}{|V(\infty)|}$ versus $\frac{\rho}{r_0}$ for $\chi_0={\sqrt{3}\over2}$.
The three lines, from bottom to top, correspond to 
$\frac{g_s^{3/4}{\bf D}}{\cN u_0^3}=$0.4, 0.8, and 10 respectively. 
We can see that for 
when the baryon density is small compared to the temperature, the thermal potential $V_T$ dominates and
the potential has a local stable minimum.
As the baryon density is increased, the relative contribution of $V_B$ becomes larger and 
eventually the potential becomes the run-away type.
}
\label{finiteTempDesdfig}
\end{center}
\end{figure}

The shape of the total potential $V=V_T+V_B$ then depends on the ratio between them, which is characterized by
\begin{eqnarray}
{V_B\over V_T} \sim {g_s\over 12\pi}
{1\over u_0^4}
\left({{\bf D}\over \cN}\right)^{{4\over3}}.
\end{eqnarray}
Physically, this suggests that the thermal effect dominates when the temperature is high (large $u_0$),
while the backreaction effect dominates when the baryon density is high
(large ${\bf D}$).\footnote{
Note that both potentials are of ${\cal O}(1/N_c)$ for fixed $g_s$
($\sim$ fixed ratio of $\lambda/N_c$).
}
Note that $\left({{\bf D}/ \cN}\right)$ can be large up to the order of $(N_c/N_f)$ where the probe flavor brane description breaks down. Therefore, the ratio $(V_B/V_T)$ may become large despite the fact that it is a positive power of $g_s$.
Recalling that $V_T$ has a local stable minimum and $V_B$ is a run-away type potential, 
we conclude that
the potential can have three possible behaviors depending on the ratio between ${\bf D}$ and $u_0$.
When $(({\bf D}/\cN)^{4/3}/u_0^4)$ is very small, $V_T$ dominates the potential and 
it has a local stable minimum.
The size of the instanton in this case is about the order of the horizon scale. 
Since observables which have less energy than the temperature have no meaning at finite temperature,
the finiteness of the instanton size may be interpreted as a thermal
effect. 
As $(({\bf D}/\cN)^{4/3}/u_0^4)$ increases, the local minimum becomes a meta-stable state and
the instanton size $\rho$ eventually decays to infinity.
As $(({\bf D}/\cN)^{4/3}/u_0^4)$ increases further,  $V_B$ dominates the potential and 
the local minimum disappears. These features are shown in Figure~\ref{finiteTempDesdfig}.
%
%\if0
%\fi
%
Thermal quantities such as the derivative of 
the free energy may change discontinuously at a critical value of $(({\bf D}/\cN)^{4/3}/u_0^4)$.
Therefore, this shows a phase transition within the CFL phase.
{For larger temperature, we have a CFL phase with the 
finite size instanton, while for
larger baryon density, we have another CFL phase with the size of the
instanton being very large. A schematic picture of the phase diagram is
shown in Fig.~\ref{figphase2}.}

As in the case of $T=0$, our analysis is valid only for small $u$ region and cannot say anything precise
about the potential at large $u$. 
From the boundary theory point of view, 
the stability at higher temperature may be understood as the thermal masses of the scalar fields
and the instability at higher density may be understood as the tachyonic masses of the scalar fields from the
chemical potential.
Since the supersymmetry is completely broken,
the potential might be lifted up by cubic and higher terms and the vevs of the squarks may take finite values.
Of course, 
these expectations are from the weak coupling analysis of the gauge theory and the strong 
coupling dynamics might change the picture.
We do not go into the detail on this point in this paper.

\if0
We would like to guess the potential behavior another way.
For zero baryon density case, the instanton potential takes its minimum at the origin $\rho=0$ for the Minkowski embeddings
while it takes the minimum at finite $\rho$ for the black hole
embeddings \cite{Apreda1}.
This suggests that the gravity force acts attractively to the instanton in the asymptotic region (large $u$) 
while it acts repulsively near the horizon.
Our analysis near the horizon shows that the backreaction effects repulsively to the potential.
\fi

\section{Discussions}
\label{sec6}

In this paper, we made some preliminary steps towards a holographic model of color-flavor locking phase, here we end with a list of interesting future directions which seem worth exploring.

The phase diagram (Fig.~\ref{figphase2}) is obtained by the total
potential $V_T + V_B$ for the instanton size modulus, but the potential
$V_B$ is valid only for a restricted region for $r$, as shown in
Sec.~\ref{sec323}. So, it is important to compute the backreaction
which is valid in all region of $r$, to explore the phase diagram
further. 

In particular for $T=0$, we have shown that there is an
instability along the direction of squark VEV in the melted meson
phase. This means that the
critical chemical potential dividing the meson and melted meson phases
may take a different value which is smaller than $\mu=m$. Our method
of treating $\tilde{N}_c$ D3-branes among $N_c$ of them separately
cannot reach the true value of the critical chemical potential in the
full phase diagram, and this deserves a further study. It is possible
that there may be no vacuum if the potential valid for all $r$ is found
and turns out to be a run-away type. See \cite{Yamada:2006rx} for 
a related discussion for R-charge chemical potential.

A related issue is a possible distinction between the two CFL Higgs
phases. We have two CFL phases, one is with finite instanton size $\rho$
while the other is with $\rho=\infty$. The former is realized mainly by
the thermal potential for $\rho$, while the latter is by a domination of
the baryon density. The symmetry breaking patterns look
similar to each other. However, we expect that, once the repulsion among
electrically charged instantons is included, the remaining symmetries
may differ. In addition, physical solitonic 
spectra in these CFL vacua may be different from each other. It would be
interesting to study vortex strings in these vacua.

The vortex strings in the CFL phase in QCD play important roles in
various physics (see Ref.~\cite{Iida:2002ev} for a partial list of related
papers), and the D-brane techniques for the CFL phase studied in 
this paper may be helpful in revealing the properties of those vortex
strings. Since the vortex strings are inside the D3-branes which are
instantons on the D7-branes, this suggests that ``vortices inside
instantons'' are possible. This is intriguing on its own in soliton
physics. 

It was described in Ref.~\cite{Schafer:1998ef} 
that in an idealistic situation the CFL
phase of QCD may be continuously connected to the hadron phase, giving a
continuous deformation of the excitation spectrum, named ``quark-hadron
continuity''.
In our case, the dynamically favored CFL phase is in the
melted meson phase, so the fluctuation spectrum is continuous, which
means that the spectral ``continuity'' doesn't make much sense. However,
in our  
${\cal N}=4$ YM theory coupled to the ${\cal N}=2$ quark hypermultiplet, 
it is known that the meson phase is continuously connected to a Higgs
phase \cite{Guralnik:2004ve}. This marginal 
deformation does not cost any energy, and the baryon number density is
kept to be zero. The instanton
size modulus is a free parameter (that is, the squark VEV is a flat
direction of the theory). In this deformation, it was shown in
Ref.~\cite{Erdmenger:2005bj} 
that the discrete fluctuation spectrum is smoothly deformed. See
Fig.~1 of Ref.~\cite{Erdmenger:2005bj}. This phenomenon is analogous to the
spectral quark-hadron continuity.

It is well-known that color-flavor locking phase in QCD closely resembles the locking between spin and orbital symmetries found in the so-called ``B-phase'' of superfluid Helium 3, the setup we consider here therefore seems to be directly applicable in realizing this in string theory. One can study various thermodynamical properites and also consider topological defects e.g. vortices and study in such phase.
Some interesting work relating D3D7 system with fermi-liquid can be found in Refs.~\cite{FermiLiquid1}.  

Finally, it would be interesting to study a possible universality of the
CFL at finite baryon density among holographic models.
In the D4/D6 system considered in Ref.~\cite{Kruczenski:2003uq}, 
the dual field theory becomes effectively a pure bosonic Yang-Mills
theory at low energy \cite{Witten:1998zw}. 
The phase structure of this system at finite temperature and baryon
density was shown to have universal properties in 
Ref.~\cite{Matsuura:2007zx}.
Therefore, it is expected that when the baryon number density increases,
the system becomes unstable and  
some of the D4-branes would be pulled onto the D6 branes.
In this case, the squarks condensation corresponds to an 
expansion of monopoles on the D6-branes, instead
of the instantons.
As mentioned in Sec.~\ref{sec3.1}, 
in the deconfinement phase, the baryon vertex is replaced by a 
flux while there is no probe brane description\cite{Seo:2008qc}.
On the other hand, in a confining phase, 
the baryon vertex does have a probe brane description, and 
the discussion in Sec.~\ref{sec3.1} at zero temperature does not 
apply to the case.
Therefore, it would be interesting to investigate the possibility of CFL in a
confinement phase.

%%%%%%%%%%%%%%%%%%%%%%%%%%%%%%%%%%%%%%%%%%%%%%%%%%%%%%%%%%%%%%%%%%%%
\acknowledgments 
We would like to thank Johanna Erdmenger, Aki Hashimoto, Deog-Ki Hong, 
Elias Kiritsis, Shin Nakamura, Hirosi Ooguri, Shigeki Sugimoto 
and Seiji Terashima for
discussions. We would also like to thank Rob Myers and David Mateos for giving many useful comments on the draft.
K.H.~thanks Kavli Institute for Theoretical Physics at UCSB
for providing an ideal environment for discussions, and thanks Aki
Hashimoto for kind hospitality to support his visit to the physics
department at University of Wisconsin. He also thanks
the Yukawa Institute for Theoretical Physics at Kyoto University, at
which this topic was discussed during the workshop YITP-W-09-04 on 
``Development of Quantum Field Theory and String Theory.''
HYC is supported in part by NSF CAREER Award No. PHY-0348093, DOE grant DE-FG-02-95ER40896, a Research Innovation Award and a Cottrell Scholar Award from Research
Corporation, and a Vilas Associate Award from the University of Wisconsin.
KH.~is partly supported by
the Japan Ministry of Education, Culture, Sports, Science and
Technology. 
The work of SM is supported in part by National Science Foundation under grant No. PHY05-51164 and Japan Society for the Promotion of Science.

\appendix

\section{Check of Consistency for the Linearized Perturbation}
\label{sec323}

To complete the analysis of Sec.~\ref{sec3.2}, 
we shall now check if this can be regarded as
a {\it small} backreaction, 
so that our perturbative treatment for solving the equations of motion
of the supergravity is guaranteed. The second term of (\ref{IIBaction})
suggests that the nonzero $F_3$ (\ref{f3back})
will again backreact the $F_5$ flux. We examine 
that this backreaction does not spoil the original flux configuration
(\ref{F5}) too much. To this end, we compare the second term of
(\ref{IIBaction}) with the $F_5$ kinetic term
\begin{eqnarray}
\frac{-1}{8\kappa_{10}^2} \int \! d^{10}x \sqrt{-g_{10}} |F_5|^2.
\label{f5kin}
\end{eqnarray}
%%%%%%%%%%%%%%%%%%%%%%%%%%%%%%%%%%%%%%%%%%%%%%%%%%%%%%%%%%%%%%%%%%%%
We are only interested in order of magnitudes. Solving (\ref{nsnseq}),
we obtain 
\begin{eqnarray}
B_{0r_6} \sim r_6^{-3} g_s^2 \alpha'^4 \bd \, .
\end{eqnarray}
Using this and (\ref{f3back}) (\ref{F5}), we evaluate the second term of
(\ref{IIBaction}) as 
\begin{eqnarray}
\frac{1}{4\kappa_{10}^2} \int F_5 \wedge B_2 \wedge F_3
\sim \int d^4x dr_6 d\Omega_5 \; r_6^{-3} g_s^2 \alpha'^4 \bd^2 \, .
\label{compare1}
\end{eqnarray}
On the other hand, the $F_5$ kinetic term (\ref{f5kin}) with the flux
solution (\ref{F5}) gives 
\begin{eqnarray}
\frac{-1}{8\kappa_{10}^2} \int \! d^{10}x \sqrt{-g_{10}} |F_5|^2
\sim
\int d^4x dr_6 d\Omega_5 \; r_6^{3} g_s^{-2} \alpha'^{-4} \, .
\label{compare2}
\end{eqnarray}
Requiring (\ref{compare1}) being much smaller than (\ref{compare2}), we
obtain 
\begin{eqnarray}
 g_s^4 \alpha'^8 \bd^2 \ll r_6^6 \, .
\label{constr1}
\end{eqnarray}
This means that, for the backreaction to the 5-form flux $F_5$ to be
small, we need to work in this region for $r_6$.

On the other hand, we made the assumption $r \ll r_0$
to simplify the source term to get (\ref{source}). 
Around the tip, we have a relation 
$r_6^2 = r^2 + y(r)^2 \sim r^2 (1 + y'(0))$, so this assupmtion
translates to the condition
$ r_6^2 (\bd^2-\bc^2)/{\bd^2} \ll r_0^2$
which is equivalent to 
\begin{eqnarray}
 r_6^6 \ll \alpha'^8 g_s^2 N_f^{-2} \bd^6/(\bd^2-\bc^2)^2 \, .
\label{constr2}
\end{eqnarray}

Therefore, in order to have a region for $r_6$ which satisfies the two
requirements (\ref{constr1}) and (\ref{constr2}), we need
\begin{eqnarray}
g_s N_f \ll \bd^2/(\bd^2-\bc^2) \, .
\label{ddc}
\end{eqnarray}
With (\ref{relm}) and (\ref{relmu}), this condition is met if we are
close to the critical chemical potential, 
\begin{eqnarray}
\mu - m \ll \mu\, , m \, .
\end{eqnarray}
Throughout this paper, we are working in this regeme.

Note that when $\bd^2-\bc^2 \ll \bd^2$ with which (\ref{ddc}) 
is satisfied, 
the D7-brane spike becomes very
narrow, and the spike can be well-approximated by fundamental strings.
This means that dilaton backreaction can be safely neglected. The
backreaction to the metric is suppressed by $1/N_c$ and also neglected.

\end{document}